\documentclass[a4paper,12pt]{article}
\pdfoutput=1 % if your are submitting a pdflatex (i.e. if you have
\usepackage[a4paper,textwidth=460.72124pt,textheight=621.0pt]{geometry}
\usepackage{cite}
\usepackage{graphicx}
\usepackage[labelfont=bf]{caption}
\usepackage{subcaption}

\usepackage[T1]{fontenc} 
\usepackage[utf8]{inputenc}

\usepackage{amssymb}
\usepackage{amsmath}
\usepackage{amsfonts}
\usepackage{mathtools}
\usepackage{etoolbox} % for simple if constructs
\usepackage{booktabs} % nicer tables

\DeclareMathAlphabet{\mathpzc}{OT1}{pzc}{m}{it}
\numberwithin{equation}{section}

\usepackage[colorlinks]{hyperref}
\hypersetup{
 citecolor=blue,
 linkcolor=blue,
 urlcolor=blue}

\usepackage[capitalize]{cleveref}

\newcommand{\la}{\left\langle}
\newcommand{\ra}{\right\rangle}
\newcommand{\lp}{\left(}
\newcommand{\rp}{\right)}
\newcommand{\lb}{\left[}
\newcommand{\rb}{\right]}

\newcommand{\ep}{\ensuremath{\epsilon}}
\newcommand{\as}{\ensuremath{\alpha_s}}
\newcommand{\asb}{\ensuremath{\alpha_{s,b}}}
\newcommand{\gs}{\ensuremath{g_s}}

\newcommand{\CA}{\ensuremath{C_A}}
\newcommand{\CF}{\ensuremath{C_F}}
\newcommand{\TF}{\ensuremath{T_F}}
\newcommand{\nf}{\ensuremath{n_f}}
\newcommand{\nl}{\ensuremath{n_l}}
\newcommand{\Nc}{\ensuremath{N_c}}

\newcommand{\Emax}{\ensuremath{E_{\mathrm{max}}}}
\renewcommand{\d}{\ensuremath{\mathrm{d}}}

\newcommand{\texparserEOL}{}

\newcommand{\LO}{{\ensuremath{\mathrm{LO}}}}
\newcommand{\NLO}{{\ensuremath{\mathrm{NLO}}}}
\newcommand{\NNLO}{{\ensuremath{\mathrm{NNLO}}}}
\newcommand{\MSbar}{\ensuremath{\overline{\textrm{MS}}}}

\newcommand{\xt}{\ensuremath{x_t}}
\newcommand{\yb}{\ensuremath{y_b}}
\newcommand{\yt}{\ensuremath{y_t}}
\newcommand{\mb}{\ensuremath{m_b}}
\newcommand{\mt}{\ensuremath{m_t}}
\newcommand{\ybbar}[1]{\ensuremath{\overline{y}_b(#1)}}
\newcommand{\mbbar}[1]{\ensuremath{\ifblank{#1}{\overline{m}_b}{\overline{m}_b(#1)}}}
\newcommand{\ybbarsq}[1]{\ensuremath{\overline{y}_b^2(#1)}}

\newcommand{\mh}{\ensuremath{M_H}}
\newcommand{\mz}{\ensuremath{M_Z}}

\newcommand{\muR}{\mu_R}
\newcommand{\order}[1]{\operatorname{\mathcal{O}}\left(#1\right)}

\newcommand{\dq}[1]{\ensuremath{[\d{}q_{#1}]}}

\newcommand{\RunDec}{{\texttt{RunDec}}}
\newcommand{\QCDLoop}{{\texttt{QCDLoop}}}
\newcommand{\HPLOG}{{\texttt{HPLOG}}}
\newcommand{\Mathematica}{{\texttt{Mathematica}}}
\newcommand{\HarmonicSums}{{\texttt{HarmonicSums}}}

\newcommand{\mev}{\ensuremath{\mathrm{MeV}}}
\newcommand{\gev}{\ensuremath{\mathrm{GeV}}}

\newcommand{\asOnPi}[1]{\ensuremath{\ifblank{#1}{\lp{\frac{\as}{\pi}}\rp}{\lp{\frac{\as(#1)}{\pi}}\rp}}}

\newcommand{\asOnFourPi}[1]{\ensuremath{\ifblank{#1}{\lp{\frac{\as}{4\pi}}\rp}{\lp{\frac{\as(#1)}{4\pi}}\rp}}}

\newcommand{\hbb}[1]{\ensuremath{H\to{b\bar{b}{#1}}}}
\newcommand{\bb}[1]{\ensuremath{{b\bar{b}{#1}}}}
\newcommand{\bbbb}{\ensuremath{{b\bar{b}b\bar{b}}}}
\newcommand{\qq}{\ensuremath{q\bar{q}}}

\newcommand{\Sep}{\ensuremath{S_\ep}}
\newcommand{\EulerGamma}{\ensuremath{\gamma_{\mathrm{E}}}}
\newcommand{\dOmega}[2]{\ensuremath{\d\Omega_{#1}^{(#2)}}}
\newcommand{\Li}{\ensuremath{\mathrm{Li}}}
\newcommand{\phsp}[1]{\ensuremath{\d\Phi_{#1}}}

\newcommand{\scprod}[2]{\ensuremath{({#1}\cdot{#2})}}

\newcommand{\DSoft}[2]{\ensuremath{\mathrm{D}{\hspace{-1pt}}\mathrm{Soft}^{(#1)}_{#2}}}
\newcommand{\Sij}[3]{\ensuremath{\mathcal{S}^{(#1)}_{#2,#3}}}

\newcommand{\Rij}[3]{\ensuremath{\mathcal{R}^{(#1)}_{#2,#3}}}
\newcommand{\Pij}[3]{\ensuremath{\mathcal{P}^{(#1),#3}_{#2}}}
\newcommand{\PijAvg}[3]{\ensuremath{\langle \mathcal{P}^{(#1)}_{#2}(#3)\rangle}}

\newcommand{\SSij}[3]{\ensuremath{\mathcal{S}^{#1}_{#2,#3}}}
\newcommand{\SijInt}[2]{\ensuremath{\mathcal{S}^{(#1)}_{#2,\mathrm{int}}}}
\newcommand{\PijInt}[2]{\ensuremath{\mathcal{P}^{(#1)}_{#2,\mathrm{int}}}}
\newcommand{\PijSoftInt}[2]{\ensuremath{\mathcal{P}^{(#1),\mathrm{soft}}_{#2,\mathrm{int}}}}
\newcommand{\DSoftInt}[3]{\ensuremath{\mathrm{D}{\hspace{-1pt}}\mathrm{Soft}^{(#1)}_{#2,\mathrm{int}}(#3)}}
\newcommand{\RInt}[1]{\ensuremath{\mathcal{R}^{(#1)}_{\mathrm{int}}}}
\newcommand{\RIntCoeff}[2]{\ensuremath{\mathcal{R}^{(#1,#2)}_{\mathrm{int}}}}

\newcommand{\MEsq}[2]{\ensuremath{|\mathcal{M}^{(#1)}_{#2}|^2}}
\newcommand{\BraAmpl}[2]{\ensuremath{\langle\ifblank{#1}{\mathcal{M}_{#2}}{\mathcal{M}^{(#1)}_{#2}}}}
\newcommand{\KetAmpl}[2]{\ensuremath{\ifblank{#1}{\mathcal{M}_{#2}}{\mathcal{M}^{(#1)}_{#2}}\rangle}}

\newcommand{\KetAmplBare}[2]{\ensuremath{\ifblank{#1}{\hat{\mathcal{M}}_{#2}}{\hat{\mathcal{M}}^{(#1)}_{#2}}\rangle}}
\newcommand{\KetCT}[2]{\ensuremath{\ifblank{#1}{\hat{\mathcal{C}}_{#2}}{\hat{\mathcal{C}}^{(#1)}_{#2}}\rangle}}

\newcommand{\Zop}[2]{\ensuremath{\ifblank{#1}{\mathbf{Z}_{#2}}{\mathbf{Z}^{(#1)}_{#2}}}}

\newcommand{\BraFinRem}[2]{\ensuremath{\ifblank{#1}{\langle\mathcal{F}_{#2}}{\langle\mathcal{F}^{(#1)}_{#2}}}}
\newcommand{\KetFinRem}[2]{\ensuremath{\ifblank{#1}{\mathcal{F}_{#2}}{\mathcal{F}^{(#1)}_{#2}}\rangle}}
\newcommand{\gamcusp}[1]{\ensuremath{\ifblank{#1}{\gamma_\mathrm{cusp}}{\gamma_\mathrm{cusp}^{(#1)}}}}
\newcommand{\gamcuspQ}[2]{\ensuremath{\ifblank{#1}{\gamma_\mathrm{cusp,Q}(#2)}{\gamma_\mathrm{cusp,Q}^{(#1)}(#2)}}}
\newcommand{\gamQ}[1]{\ensuremath{\ifblank{#1}{\gamma_Q}{\gamma_Q^{(#1)}}}}
\newcommand{\gamg}[1]{\ensuremath{\ifblank{#1}{\gamma_g}{\gamma_g^{(#1)}}}}

\newcommand{\ident}{\ensuremath{I}}
\newcommand{\soft}[1]{\ensuremath{S_{#1}}}
\newcommand{\coll}[1]{\ensuremath{C_{#1}}}
\newcommand{\dsoft}[1]{\ensuremath{S_{#1}}}

\newcommand{\FLM}[2][]{\ensuremath{{#1}F_{LM}(#2)}}

\newcommand{\FLMInt}[2][]{\ensuremath{\langle{#1}\FLM{#2}\rangle}}
\newcommand{\FLV}[2][]{\ensuremath{{#1}F_{LV}(#2)}}

\newcommand{\FLVfin}[2][]{\ensuremath{{#1}F_{LV}^{\mathrm{fin}}(#2)}}
\newcommand{\FLVfinInt}[2][]{\ensuremath{\langle{#1}F_{LV}^{\mathrm{fin}}(#2)\rangle}}

\newcommand{\obs}[1]{\ensuremath{\ifblank{#1}{\mathcal{F}_{\textrm{kin}}}{\mathcal{F}_{\textrm{kin}}(#1)}}}
\newcommand{\dGam}[3]{\ensuremath{ {\d\Gamma}_{#1}^{#2}{ \ifblank{#3}{}{(#3)} } }}
\newcommand{\dGamInt}[3]{\ensuremath{\langle {\d\Gamma}_{#1}^{#2}{ \ifblank{#3}{}{(#3)} } \rangle}}

\newcommand{\R}{\ensuremath{\mathrm{R}}}
\newcommand{\V}{\ensuremath{\mathrm{V}}}
\newcommand{\RR}{\ensuremath{\mathrm{RR}}}
\newcommand{\RV}{\ensuremath{\mathrm{RV}}}
\newcommand{\VV}{\ensuremath{\mathrm{VV}}}
\newcommand{\F}{\ensuremath{\mathrm{F}}}
\newcommand{\U}{\ensuremath{\mathrm{U}}}
\newcommand{\FR}{\ensuremath{\mathrm{FR}}}
\newcommand{\SU}{\ensuremath{\mathrm{SU}}}
\newcommand{\DU}{\ensuremath{\mathrm{DU}}}

\newcommand{\partcoeff}[3]{\ensuremath{\ifblank{#3}{\mathcal{C}_{#1}^{#2}}{\mathcal{C}_{#1}^{#2,(#3)}}}}

\newcommand{\gmn}{\ensuremath{g^{\mu\nu}}}
\newcommand{\ktmu}{\ensuremath{k_{\perp}^{\mu}}}
\newcommand{\ktnu}{\ensuremath{k_{\perp}^{\nu}}}
\newcommand{\ktsq}{\ensuremath{k_{\perp}^{\,2}}}

\newcommand{\ycut}{\ensuremath{y_{\mathrm{cut}}}}

\renewcommand{\Re}{\operatorname{Re}}

\begin{document}

%>>> front page
\pagenumbering{roman}
\pagestyle{empty}
\vspace{-5.0cm}
\begin{flushright}
  P3H-19-047, TTP19-041
\end{flushright}

\vspace{2.0cm}
\begin{center}
  {\large \textbf{
    Higgs decay into massive $b$-quarks at NNLO QCD \\ in the nested soft-collinear subtraction scheme
  }}\\
\end{center}

\vspace{0.5cm}
\begin{center}
  Arnd Behring$^{a}$,
  Wojciech Bizo\'n$^{a,b}$.
  \\
  \vspace{0.3cm}
  {\textit{
    {}$^{a}$ Institut f{\"u}r Theoretische Teilchenphysik (TTP), KIT, 76128 Karlsruhe, Germany\\
    {}$^{b}$ Institut f{\"u}r Kernphysik (IKP), KIT, 76344 Eggenstein-Leopoldshafen, Germany\\
  }}
\end{center}

\vspace{1.3cm}
\begin{center}
{\large \textbf{Abstract} }
\end{center} {
  We present a fully differential description of a decay of a scalar
  Higgs boson into massive $b$-quarks valid at next-to-next-to-leading
  order (NNLO) in perturbative quantum chromodynamics (QCD).
  We work within the nested soft-collinear subtraction scheme
  extended to accommodate massive partons.
  We include the loop-induced contribution involving a Higgs coupling to a top quark.
  We test our calculation against results existing in the literature,
  comparing the predictions for the total decay width and jet rates.
}
\clearpage

%>>> table of contents
{
  \hypersetup{linkcolor=black}
  \tableofcontents
}
\clearpage

%>>> text
\pagestyle{plain}
\pagenumbering{arabic}
\allowdisplaybreaks

%-------------------------
\section{Introduction}
\label{sec:intro}
%%%
After the Higgs boson discovery by the ATLAS and CMS collaborations in
2012, the study of Higgs boson properties has become one of the major
research avenues in particle physics.
Since the mass of the Higgs boson has already been precisely
measured~\cite{Aad:2015zhl}, all couplings between the Higgs boson and
other Standard Model (SM) particles can be accurately predicted.
Nevertheless, these couplings can be modified by New Physics that lies
beyond the SM. Hence, actual measurements of those couplings can provide
important constraints on many extensions of the SM.

%%%
The Higgs boson decay into a pair of $b$-quarks is the most common
decay channel of the Higgs boson and it is essential to many New
Physics searches.
Indeed, it plays a particularly important role when considering rare
Higgs boson production modes, which benefit from the large \hbb{}
branching fraction.
Although such measurements are often very challenging, due to
overwhelming QCD backgrounds, the \hbb{} decay has already been
observed by both ATLAS and
CMS~\cite{Aaboud:2018zhk,Sirunyan:2018kst}.
The upcoming years of data taking at the Large Hadron Collider (LHC)
will allow for further exploration and use of this Higgs boson decay
channel.

%%%
In order to fully utilise the data collected at the LHC, a good
theoretical understanding of the \hbb{} process is required.
The next-to-leading order (NLO) QCD corrections have been available for
a long time~\cite{Braaten:1980yq,Sakai:1980fa,Janot:1989jf,%
Drees:1990dq,Kataev:1992fe}.
Currently, corrections to the total decay width are known up to
$\mathcal{O}(\as^4)$ in the limit of massless
$b$-quarks~\cite{Baikov:2005rw}.
Mass effects at $\mathcal{O}(\as^2)$ have been estimated using a large
momentum expansion~\cite{Harlander:1997xa}.
Furthermore, the impact of a separate class of corrections from
diagrams arising at $\mathcal{O}(\as^2)$ which involve the top-quark
Yukawa coupling has been calculated in the limit of a large top-quark
mass~\cite{Larin:1995sq,Chetyrkin:1995pd} as well as for general
values of the masses~\cite{Primo:2018zby}.
Recently, a set of two-loop master integrals required for mixed
QCD-electroweak corrections has been computed~\cite{Chaubey:2019lum}.
In the limit of massless $b$-quarks a number of fully differential
next-to-next-to-leading order (NNLO) calculations of the \hbb{} decay
have been presented~\cite{Anastasiou:2011qx,DelDuca:2015zqa,%
Caola:2017xuq,Gauld:2019yng} with first N$^3$LO QCD results appearing
recently~\cite{Mondini:2019gid}.
The $b$-quark mass effects for differential observables have been
studied at NNLO QCD in Ref.~\cite{Bernreuther:2018ynm}.

In this paper, we present an independent calculation of the $b$-quark
mass effects in the \hbb{} decay at NNLO QCD.
Although we believe such a calculation is interesting in its own right
and serves as a useful check of the results presented in
Ref.~\cite{Bernreuther:2018ynm}, it is also an essential step towards
studying mass effects in associated Higgs production with a vector
boson, $pp \to HV \to \bb{}V$.
NNLO QCD corrections to this process have already been studied in the
limit of massless
$b$-quarks~\cite{Ferrera:2017zex,Caola:2017xuq,Gauld:2019yng}, and
large effects, related to radiative corrections, have been reported.
The impact of the $b$-quark mass may be sizeable in certain regions of
the phase space.
Higher-order effects in that process have also been investigated using
parton showers~\cite{Astill:2018ivh,Alioli:2019qzz,Granata:2017iod}.
Another contribution to the Higgs decay width at NNLO is mediated by
top-quark loops. In the context of differential distributions, it has
been discussed in Ref.~\cite{Caola:2017xuq} and subsequently
investigated in Ref.~\cite{Primo:2018zby}.
A consistent treatment of these contributions requires keeping the
$b$-quark mass finite~\cite{Caola:2017xuq}.

%%%
We work within the nested soft-collinear subtraction
scheme~\cite{Caola:2017dug,Caola:2017xuq,Caola:2019nzf,Caola:2019pfz},
which is an extension of the original sector-improved residue
subtraction
scheme~\cite{Czakon:2010td,Czakon:2011ve,Czakon:2014oma,Czakon:2019tmo}.
To incorporate the $b$-quark masses into the calculation, we rely on
the treatment of massive particles outlined in
Ref.~\cite{Czakon:2014oma}.

%%%
The paper is organised as follows.
In \cref{sec:general} we introduce the notation and discuss the main
steps of the subtraction scheme. We also describe the infrared (IR)
poles appearing in virtual amplitudes and touch upon the relation
between the pole and \MSbar{} Yukawa couplings.
In \cref{sec:nlo} and \cref{sec:nnlo} we review the NLO QCD
calculation of the \hbb{} decay and present our calculation of the
NNLO corrections, including the treatment of the top-quark induced
corrections.
Finally, in \cref{sec:results}, we thoroughly test our predictions
against results available in the literature. We summarise our findings
in \cref{sec:summary}.
The appendices contain a number of expressions used throughout our
calculation.

%-------------------------
\section{General considerations}
\label{sec:general}
We are interested in decays of a scalar Higgs boson into a pair of
$b$-quarks.  Our goal is to treat $b$-quarks as massive particles
throughout the calculation and achieve NNLO accuracy in perturbative QCD
while working within the nested soft-collinear subtraction
scheme~\cite{Caola:2017dug,Caola:2017xuq,Caola:2019nzf,Caola:2019pfz}.

\subsection{Notation}
We start with a short introduction that will set the stage for our
calculation. We consider the Higgs boson decaying into a pair of
$b$-quarks
\begin{align}
  H(q_1)
  \longrightarrow
  b(q_2) + \bar{b}(q_3)\,,
\end{align}
with $q_1^{\,2} = \mh^{\,2}$ and $q_2^{\,2} = q_3^{\,2} = \mb^{\,2}$.
The leading order (LO) partial decay width of the Higgs boson into a
pair of $b$-quarks is
\begin{align}
  \Gamma_{\LO}
  ={}&
       \frac{1}{2\mh}
       \int\d\Phi_{\bb{}}(q_1)
       \MEsq{0}{\bb{}}\,,
\end{align}
where $\d\Phi_{\bb{}}(q_1)$ is the two-particle phase-space volume
element for the production of two $b$-quarks with total momentum
$q_1$, and $\MEsq{0}{\bb{}}$ is the squared tree-level matrix element.
For brevity, we list only final-state particles in the sub- and
superscripts, since the initial state is always an on-shell Higgs
boson at rest. Upon integration we obtain
\begin{align}
  \Gamma_{\LO}
  ={}&
       \frac{\Nc}{16\pi}
       \yb^2 \mh \beta^3
       \,,
\end{align}
where $\Nc=3$ is the number of colours,
$\beta = \sqrt{1 - 4\mb^2/\mh^2}$ and $\yb$ stands for the $b$-quark
Yukawa coupling, $\yb = \mb(2\sqrt{2}G_F)^{1/2}$.

%% outline
The Higgs boson decay width into $b$-quarks receives radiative
corrections that can be systematically calculated in perturbative
QCD. We use the following notation for the perturbative expansion of
the width
\begin{align}
  \label{eq:width-expansion}
  \Gamma^{\bb{}}
  ={}&
       \Gamma_{\LO}^{\bb{}}
       \lb{
       1
       + \asOnPi{} \gamma_{1}^{\bb{}}
       + \asOnPi{}^2 \gamma_{2}^{\bb{}}
       + \mathcal{O}(\as^3)
       }\rb\,.
\end{align}
In \cref{eq:width-expansion}, $\as$ is the \MSbar{} QCD strong
coupling constant defined in a theory with $\nf = \nl + 1$ quark
flavours, where $\nl$ is the number of massless flavours.

%% abbreviations
It is useful to define a shorthand notation that denotes an integral
over the Lorentz-invariant phase space of particles involved in a
particular (sub)process.
Similar to the notation in Ref.~\cite{Caola:2017dug}, we define
\begin{align}
  \label{eq:FLM_def}
  \FLM{\bb{X}} = \d\Phi_{\bb{X}}(q_1) \MEsq{0}{\bb{X}} \obs{\bb{X}}\,,
\end{align}
where \bb{X} denotes the constituents of the final state of a
considered subprocess, $\d\Phi_{\bb{X}}(q_1)$ and $\MEsq{0}{\bb{X}}$
stand for the Lorentz-invariant phase-space measure and the squared
amplitude of the \hbb{X} process, respectively.
The momentum $q_1$ refers to the initial state Higgs boson, while
\obs{} is an infrared-safe observable that depends on the kinematical
configuration of the particles involved in the process.
We will use the notation
\begin{align}
  \langle A \rangle &= \int \d\Phi A
\end{align}
to denote the integration of some quantity $A$ over the phase space,
$\d\Phi$.

%%%
\subsection{Outline of the subtraction scheme}
\label{sec:philosophy}
One of the challenges in higher-order QCD calculations is the
appearance of infrared singularities when massless particles become
soft or collinear.
Dimensional regularisation~\cite{tHooft:1972tcz,
  Ashmore:1972uj,Cicuta:1972jf,Bollini:1972ui,Marciano:1974tv} can be
used to regulate these singularities which show up as poles in the
dimensional regularisation parameter $\ep = (4-d)/2$ in both real and
virtual amplitudes.
For an infrared-safe observable these poles cancel between real and
virtual corrections and collinear factorisation contributions once the
loop and phase-space integrals over the singular regions are
performed~\cite{Kinoshita:1962ur,Lee:1964is}.
The observables depend on the momenta of the real emission partons so
that numerical integration of the phase-space integrals is desirable
from the standpoint of flexibility and often also required due to the
complexity of the observable, which may involve, e.g., complicated
kinematic constraints and jet algorithms.

It follows from the factorisation theorems of QCD that the integrand of
the cross-section, i.e., the combination of squared matrix elements and
phase-space measure, scales as $E_i^{-1+a \ep}\d{E_i}$ in soft limits
and as $\theta_{ij}^{-1+b \ep}\d\theta_{ij}$ in collinear limits of
massless partons, where $E_i$ is the energy of the soft parton,
$\theta_{ij}$ is the angle between the collinear partons and $a,b \in
\mathbb{R}$. Thus, $\ep$ acts as a regulator for logarithmic divergences
in the real-emission phase-space integrals. For the numerical
integration it is necessary to explicitly extract and cancel the poles
in $\ep^{-1}$ and to regulate the integrals in such a way that the
expansion in $\ep$ can be performed at the integrand level.

A number of methods have been developed to accomplish that. Here, we
follow the nested soft-collinear subtraction
scheme~\cite{Caola:2017dug,Caola:2017xuq,Caola:2019nzf,Caola:2019pfz}
which is closely related to the sector-improved residue subtraction
scheme~\cite{Czakon:2010td,Czakon:2011ve,Czakon:2014oma,Czakon:2019tmo}.
As with all subtraction schemes, the general idea is to introduce
subtraction terms for each singular limit which regulate the integrand
and to add back these subtraction terms integrated over the unresolved
phase space. Schematically, for an integral of a function $F$ over the
phase space, we write
\begin{align}
  \langle F \rangle &= \langle F - OF \rangle  + \langle OF \rangle
  \label{eq:subtraction-master-formula}
  \,,
\end{align}
where $O$ is an operator which extracts the asymptotic behaviour of
$F$ and the phase space in a singular limit. The term
$\langle OF \rangle$, which is integrated over the unresolved phase
space in $d$ dimensions, then carries explicit poles in $\ep^{-1}$,
while the regulated term $\langle F - OF \rangle$ is expanded in $\ep$
at the integrand level.  We apply \cref{eq:subtraction-master-formula}
recursively to regulate all singular limits.

The nested soft-collinear subtraction scheme consists of the following
steps.
\begin{enumerate}
  \item Introduce subtraction terms for the soft limits. At NNLO this
        involves up to two single-soft limits and the double-soft limit.
  \item Introduce a partition of unity for the phase space that isolates
        the collinear limits, $1 = \sum_{\{p\}} w_{\{p\}}$, where the
        sum runs over the sets of partons that can produce collinear
        singularities and the functions $w_{\{p\}}$ go to zero whenever
        two partons which are not in ${p}$ become collinear, thereby
        regulating integrand in that limit.
  \item In each collinear partition, choose a suitable phase-space
        parametrisation in terms of angles and energies of the partons
        that can become unresolved.
  \item Use sector decomposition~\cite{Binoth:2000ps,Anastasiou:2003gr,%
        Binoth:2004jv} to map all singularities to the boundaries of
        the region of integration so that
        the singularities can be easily extracted upon using 
        \cref{eq:subtraction-master-formula}. In order to generate
        the limits of the matrix elements, we use the standard QCD
        factorisation formulae for the soft and collinear limits.
        All necessary expressions up to NNLO can be found, e.g., in the
        appendices of Ref.~\cite{Czakon:2014oma}.
\end{enumerate}
The $\hbb{}$ process with massive $b$-quarks is particularly simple in
this context since there are no triple- or double-collinear limits
that involve $b$-quarks so that step 2 can be avoided. Moreover, by
choosing an appropriate phase-space parametrisation in step 3, the
sector decomposition of step 4 only yields a single collinear
subsector.

The calculation will be subdivided into pieces so that the
cancellation of $\ep^{-1}$ poles can be shown without making reference
to the explicit form of the matrix elements. In particular, we will
organise the contributions into sets according to the multiplicity of
the resolved final state and the loop order of the matrix elements.
By combining this with a suitable phase-space parametrisation which
decouples the integrations over the unresolved and resolved parts of
phase space, we demonstrate pole cancellation for each phase-space
point of the resolved configuration separately.
In \cref{sec:nlo,sec:nnlo} we explain the application of
this scheme to the $\hbb{}$ process in greater detail.

\subsection{IR poles of virtual amplitudes}
\label{sec:virtIR}
The one- and two-loop virtual amplitudes that we encounter in an NNLO
calculation feature ultraviolet (UV) divergences that can be removed
using a suitable renormalisation procedure. We employ a hybrid scheme in
which we renormalise quark and gluon fields, the quark masses and the
Yukawa coupling in an on-shell scheme, while we use the \MSbar{} scheme
for the strong coupling.  The details of our renormalisation choice are
described in \cref{sec:renorm}.

At this point, the renormalised amplitudes are free of UV divergences.
Nevertheless, they still contain poles in $\ep^{-1}$ which are of IR
origin. These poles can be predicted from general considerations
\cite{Catani:1998bh,Catani:2000ef,Aybat:2006mz,Becher:2009kw,%
Czakon:2009zw,Mitov:2009sv,Ferroglia:2009ii,Mitov:2010xw,%
Becher:2009cu}. They factorise in the form
\begin{align}
  \label{eq:Zop-def}
  |\KetAmpl{}{} &= \Zop{}{} |\KetFinRem{}{}
  \,,
\end{align}
where $|\KetAmpl{}{}$ is the UV-renormalised amplitude, $\Zop{}{}$ is
an operator in colour space which contains poles in $\ep^{-1}$ and
$|\KetFinRem{}{}$ is a finite remainder which does not contain any
poles. Expanding all pieces in the strong coupling, we find
\begin{align}
  \label{eq:Z_op_expansion}
  |\KetAmpl{}{}
    ={}& |\KetAmpl{0}{}
         +\frac{\as}{4\pi} \left(
           \Zop{1}{} |\KetAmpl{0}{}
           +|\KetFinRem{1}{}
         \right)
  \notag \\ &
         +\left(\frac{\as}{4\pi}\right)^2 \left(
           \Zop{2}{} |\KetAmpl{0}{}
           +\Zop{1}{} |\KetFinRem{1}{}
           +|\KetFinRem{2}{}
         \right)
  \,.
\end{align}
Note that this notation leaves all powers of the strong coupling
related to real emissions implicit inside the amplitudes.
On the one hand, we can use \cref{eq:Z_op_expansion} as a prediction
in order to check the $\ep^{-1}$ poles of the UV-renormalised
amplitudes. On the other hand, we can also use
\cref{eq:Z_op_expansion} to define the finite remainders, i.e.
\begin{align}
  \label{eq:AmplFinRem_def}
  |\KetFinRem{0}{} &= |\KetAmpl{0}{}
  \,, \\
  |\KetFinRem{1}{} &= |\KetAmpl{1}{} - \Zop{1}{} |\KetAmpl{0}{}
  \,, \\
  |\KetFinRem{2}{} &= |\KetAmpl{2}{} - \Zop{1}{} |\KetFinRem{1}{}
                      -\Zop{2}{} |\KetAmpl{0}{}
  \,,
\end{align}
and express all formulae in terms of these. This is useful for showing
pole cancellation since it allows us to make the pole terms explicit
without specifying the matrix elements that they multiply. The pole
terms are multiplied by lower order quantities, as expected. Note that
we only include the $\ep^{-1}$ poles in the definition of the
$\Zop{}{}$ operator, cf. Refs.~\cite{Becher:2009kw,Becher:2009cu}.

%% H->bb poles
In general, $\Zop{}{}$ is an operator acting on vectors in colour
space, which expresses non-trivial correlations between different
colour configurations~\cite{Catani:1996vz,Becher:2009cu}.
However, in our case, the coefficients can be expressed in terms of
simple colour factors since we only require virtual amplitudes with up
to three coloured particles (\hbb{} and \hbb{g}). The expansion
coefficients $\Zop{k}{\bb{}}$ for the \hbb{} process are given by
%%%texparser:start:Zop%%%
\begin{align}
  \label{eq:hbb_Zop1}
  %%%texparser:LHS:Zop1bb%%%
  \Zop{1}{\bb{}}
  ={}&
       \frac{1}{2\ep} (\CF \gamcuspQ{0}{v_{23}} + 2\gamQ{0})
       \,, \\
  \label{eq:hbb_Zop2}
  %%%texparser:LHS:Zop2bb%%%
  \Zop{2}{\bb{}}
  ={}&
       \frac{1}{8\ep^2}
       (\CF \gamcuspQ{0}{v_{23}} + 2 \gamQ{0})
       (\CF \gamcuspQ{0}{v_{23}} + 2 \gamQ{0} - 2 \beta_0(\nl))
       \notag
  \\ &
       +\frac{1}{4\ep} (\CF \gamcuspQ{1}{v_{23}} + 2\gamQ{1})
       +\frac{1}{2\ep} (\CF \gamcuspQ{0}{v_{23}} + 2\gamQ{0})
         \beta_{0,Q} \ln\lp{\frac{\muR^2}{\mb^2}}\rp
       \,,
\end{align}
%%%texparser:stop%%%
where $\gamcusp{i}$ and $\gamcuspQ{0}{v_{23}}$ are the massless and
massive cusp anomalous dimensions,
$v_{23} = \sqrt{1 - \mb^4/\scprod{q_2}{q_3}^2}$, $\beta_0(\nl)$ is the
zeroth-order coefficient of the QCD $\beta$-function with $\nl$ massless
flavours, $\beta_0(\nl) = \frac{11}{3} \CA - \frac{4}{3} \TF \nl$ and
$\beta_{0,Q} = -\frac{4}{3}\TF$. The $\gamQ{i}$ denote the expansion
coefficients of the anomalous dimensions of the massive quark and $\muR$
is the renormalisation scale. We collect the necessary formulae in
\cref{sec:Zcoeff}.
For the $\hbb{g}$ process, where we only need the one-loop amplitude,
we find
%%%texparser:start:Zop%%%
\begin{align}
  \label{eq:hbbg_Zop1}
  %%%texparser:LHS:Zop1bbg%%%
  \Zop{1}{\bb{g}}
  ={}&
       \frac{1}{4\ep^2} (-\CA \gamcusp{0})
       +\frac{1}{2\ep} \left[
       \gamg{0} + 2 \gamQ{0}
       +\left(\CF-\frac{\CA}{2}\right) \gamcuspQ{0}{v_{23}}
       \right.
       \notag \\
     & \left.
       -\frac{\CA}{2} \gamcusp{0} \left(
       \ln\left(\frac{\mb \muR}{2 \scprod{q_2}{q_4}}\right)
       +\ln\left(\frac{\mb \muR}{2 \scprod{q_3}{q_4}}\right)
       + 2 i \pi
       \right)
       \right]
       \,.
\end{align}
%%%texparser:stop%%%
The gluon anomalous dimension $\gamg{0}$ is also given in
\cref{sec:Zcoeff}.

\subsection{Phase-space parametrisation}
\label{sec:phsp_param}
In this section we outline the parametrisation of the real-emission
phase space that we employ throughout the calculation.
NNLO corrections to Higgs decays involve contributions with up to two
real emissions accompanying the Born process. In our case, the Born
process consists of the Higgs boson decaying into massive $b$-quarks
(\hbb{}) and we include final states with one additional gluon
(\hbb{g}) as well as two additional massless partons (\hbb{gg} or
\hbb{\qq{}})\footnote{The \hbb{\bb{}} contribution is finite. Thus,
  it can be integrated using standard techniques and we do not
  discuss it here.}.

The guiding principle behind the construction outlined in this section
is, first, to explicitly parametrise the energies and angles that are
responsible for the soft and collinear singularities and, second, to
decouple the real-emission phase space from the phase space of the
reduced process once a parton becomes unresolved.
We note that we work in the Higgs boson rest frame throughout the
paper.

%%%
The phase-space measure for an emission of a single massless parton
in $d=4-2\ep$ space-time dimensions reads
\begin{align}
  \label{eq:LISP_parton}
  \dq{i}
  ={}&
       (\muR^2)^{\ep} \Sep
       \frac{\d^{d-1}q_i}{(2\pi)^{d-1} (2 E_i)}
       \,,
\end{align}
where we denote the parton momentum by $q_i$ and its energy by $E_i$.
Note that we also include a global factor $(\muR^2)^{\ep} \Sep$ that
originates from the strong coupling renormalisation, see discussion
below \cref{eq:renoHbbg}.  We do not introduce an upper bound on the
energy of the emitted gluon since it naturally appears due to the
energy-momentum conserving $\delta$-function once the measure in
\cref{eq:LISP_parton} is considered as a part of a specific process.

%%% H -> bbg
\paragraph{Single-emission phase space}
We start with the process
\begin{align}
  H(q_1)
  \longrightarrow
  b(q_2) + \bar{b}(q_3)
  + g(q_4)
  \,,
\end{align}
and discuss its phase-space parametrisation.

The phase-space measure reads
\begin{align}
  \int \d\Phi_{\bb{g}}(q_1)
  &={}
    \int \dq{4}
    \int \d\Phi_{\bb{}}(q_1 - q_{4})
    \nonumber\\
  &={}
    (\muR^2)^{\ep} \Sep
    \int \frac{\d^{d-2}\hat{q}_4}{2(2\pi)^{d-1}}
    \int \d E_4 (E_4)^{d-3}
    \int \d\Phi_{\bb{}}(q_1 - q_{4}),
\end{align}
where $\d\Phi_{\bb{}}(Q)$ stands for the Born phase space of the two
$b$-quarks with total momentum $Q$, and $\hat{q}_4$ determines the
direction of the gluon momentum,
\begin{align}
  \hat{q}_4^{\mu} ={}& q_4^{\mu} / E_4 \,.
\end{align}
We further parametrise the gluon energy as
\begin{align}
  E_4 &= \Emax \, \xi_1,
\end{align}
with $\Emax = \tfrac{1}{2}\beta^2\mh$ and
$\beta = \sqrt{ 1 - 4\mb^2/\mh^2}$. This finally leads us to
\begin{align}
  \label{eq:phsp_hbbg}
  \int \d\Phi_{\bb{g}}(q_1)
  ={}&
    2^{1-2\ep} \Emax^{2}
    \lp{\frac{ \muR^2 }{ \Emax^2}}\rp^\ep
    \Sep
    \int \frac{\dOmega{4}{2-2\ep}}{2 (2\pi)^{3-2\ep}}
  \nonumber \\ &
    \times
    \int_0^1 \frac{\d\eta}{(\eta(1-\eta))^\ep}
    \int_0^1 \d \xi_1
    \xi_1^{1-2\ep}
    \int \d\Phi_{\bb{}}(q_1-q_{4})
    \,.
\end{align}
The $\d\Omega_4$ element denotes the angular integral over the
direction of $\hat{q}_4$. The remaining angular integrals can be
performed using
\begin{align}
  \label{eq:dOmega_int}
  \int \d\Omega^{(a)} &={} \frac{ 2\pi^{a/2} }{\Gamma(a/2)} \,.
\end{align}

For a single gluon emission, the only unresolved limit is the
single-soft one, i.e. $\xi_1 \to 0$.  Obviously, this removes $q_4$
from the overall momentum conservation and the integration over the
unresolved phase space of $q_4$ decouples from the Born phase space.
Thus, in that limit we just replace $\int\d\Phi_{\bb{}}(q_1-q_4)$ by
$\int\d\Phi_{\bb{}}(q_1)$ in \cref{eq:phsp_hbbg}.

%%% H -> bbgg
\paragraph{Double-emission phase space}
We now focus on the parametrisation of the \hbb{gg} phase space. Note
that, since we consider $b$-quarks to be massive, there are no
singularities associated with kinematic configurations where gluons
become collinear to $b$-quarks. Therefore, we do not need to partition
the phase space into subsectors, which are usually necessary to
disentangle the collinear singularities. Instead, we work with a
global parametrisation.
In this section we focus on the two-gluon emission case since the
parametrisation of the $\qq$ emission phase space is nearly identical.
We comment on the differences where necessary.

%%%
We consider the process
\begin{align}
  H(q_1)
  \longrightarrow
  b(q_2) + \bar{b}(q_3)
  + g(q_4) + g(q_5)
  \,.
\end{align}
We denote the sum of the gluon momenta by $q_{45} = q_4 + q_5$. The
phase space measure then reads
\begin{align}
  \int \d\Phi_{\bb{gg}}(q_1)
  ={}&
    \int \dq{4}
    \int \dq{5}
    \int \d\Phi_{\bb{}}(q_1 - q_{45})
    \nonumber\\
  ={}&
    (\muR^2)^{2\ep} \Sep^2
    \int \frac{\d^{d-2}\hat{q}_4}{2(2\pi)^{d-1}}
    \int \frac{\d^{d-2}\hat{q}_5}{2(2\pi)^{d-1}}
    \nonumber\\
  &\hphantom{(\muR^2)^{2\ep}}
    \times
    \int \d E_4 (E_4)^{d-3}
    \int \d E_5 (E_5)^{d-3}
    \int \d\Phi_{\bb{}}(q_1 - q_{45}),
\end{align}
where the vectors $\hat{q}_{4}$ and $\hat{q}_5$ determine the
directions of the two gluons and the limits of the energy integrals
are so that the whole phase space is covered.

It is convenient to introduce an energy ordering among the gluons by
partitioning the phase space via
\begin{align}
  1 &={} \Theta( E_4 - E_5 ) + \Theta( E_5 - E_4 )\,
\end{align}
which leads to the split
\begin{align}
  \label{eq:phsp_split}
  \int \d\Phi_{\bb{gg}}(q_1)
  ={}&
       \int \d\Phi_{\bb{gg}}^{E_4 > E_5}(q_1)
       + \int \d\Phi_{\bb{gg}}^{E_5 > E_4}(q_1)\,.
\end{align}
Throughout the article, we describe calculations only for the region
with $E_4 > E_5$; the other region can easily be covered by performing
the same steps with the gluon momenta swapped,
$q_4 \leftrightarrow q_5$.\footnote{Note that thanks to the symmetry
  between the gluons we have
  $\int \d\Phi_{\bb{gg}}(q_1)=2 \int \d\Phi_{\bb{gg}}^{E_4 >
    E_5}(q_1)$. However when considering a $\qq$ emission such a
  simplification may only be used if the symmetry
  $q\leftrightarrow{\bar{q}}$ also holds for the observable under
  consideration. Otherwise the two parts of the phase space,
  introduced in \cref{eq:phsp_split}, have to be considered
  separately. } Hence, we parametrise the gluon energies as
\cite{Czakon:2019tmo}
\begin{align}
  \label{eq:E4_E5_param}
  E_4 &={} E_{45,\mathrm{max}} \, \xi_1 \lp 1-\frac{\xi_2}{2} \rp
  \,, &
  E_5 &={} E_{45,\mathrm{max}} \, \xi_1 \frac{\xi_2}{2}
  \,,
\end{align}
where $E_{45,\mathrm{max}}$ is to be chosen such that the integration
ranges $\xi_1 \in [0,1]$ and $\xi_2 \in [0,1]$ span the whole phase
space. Using momentum conservation and considering a configuration
where the two $b$-quarks are produced at threshold, we obtain
\begin{align}
  \label{eq:E45max}
  E_{45,\mathrm{max}}
  &={}
    \frac{ \mh \beta^2 }{ 1 + \sqrt{ 1 - \beta^2 \bar{q}_{45}^2 } }
    \,,
\end{align}
where $\bar{q}_{45} = q_{45}/q^0_{45}$, which is only a light-like momentum
when $q_5$ is soft or $q_4$ and $q_5$ are collinear. In this way, we
effectively parametrise the sum of the gluon energies
($\xi_1=E_{45}/E_{45,\mathrm{max}}$) and their ratio ($\xi_2 = 2
E_5/E_{45}$).

We also explicitly parametrise the angle $\theta_{45}$ between the two
gluons as follows
\begin{align}
  \label{eq:eta}
  \eta &= \frac{1}{2} (1-\cos\theta_{45})
  \,.
\end{align}

The first step in the phase-space construction is to choose a
direction for $\hat{q}_4$. Here, we explicitly parametrise angle
$\theta_{4}$ between the emission and the $\hat{z}$-axis.
Next, we fix the direction of $\hat{q}_5$ relative to $\hat{q}_4$ using
the angle $\theta_{45}$ between them and the angle $\phi$ which is the
azimuthal angle of $\hat{q}_5$ around the direction of $\hat{q}_4$.
Given the directions $\hat{q}_4$ and $\hat{q}_5$ as well as $\xi_2$,
we calculate the vector $\bar{q}_{45}$ via
\begin{align}
  \label{eq:q45hat}
  \bar{q}_{45} &= \left(1-\frac{\xi_2}{2}\right) \hat{q}_4
                  +\frac{\xi_2}{2} \hat{q}_5
  \,.
\end{align}
This is sufficient to use \cref{eq:E45max} in order to calculate the
upper bound on the energy and from that also the individual energies
$E_4$ and $E_5$, which fully determines $q_4$ and $q_5$.
Then we generate a Born phase-space configuration with invariant mass
$Q^2 = (q_1 - q_{45})^2$ in its rest frame; this is a back-to-back
configuration of the two $b$-quarks.
Finally, we boost the Born configuration to have total momentum
$Q = q_1 - q_{45}$ in the Higgs rest frame in order to restore
momentum conservation.
The corresponding phase-space measure reads
\begin{align}
  \label{eq:phsp_hbbgg}
  \MoveEqLeft{\int \d\Phi_{\bb{gg}}^{E_4 > E_5}(q_1)
  ={}
  2^{2\ep-2}
  (\muR^2)^{2\ep} \Sep^2
  }&
       \nonumber\\
     &
       \times
       \int \frac{\dOmega{4}{2-2\ep}}{2 (2\pi)^{3-2\ep}}
       \int_0^\pi \d\theta_4 (\sin\theta_4)^{1-2\ep}
       \int \frac{\dOmega{5}{1-2\ep}}{2 (2\pi)^{3-2\ep}}
       \int_0^1 \frac{ \d\eta }{ \eta^\ep (1-\eta)^\ep }
       \int_0^\pi \d\phi (\sin\phi)^{-2\ep}
       \nonumber\\
     &
       \times
       \int_0^1 \d \xi_1 \int_0^1 \d \xi_2
       E_{45,\mathrm{max}}^{4-4\ep}
       \xi_1^{3-4\ep} \xi_2^{1-2\ep} (2-\xi_2)^{1-2\ep}
       \int \d\Phi_{\bb{}}(q_1-q_{45})
       \,.
\end{align}
The unparametrised angles in $\d\Omega_4$ and $\d\Omega_5$ can be
integrated in the end using \cref{eq:dOmega_int}.

The parametrisation shown in \cref{eq:phsp_hbbgg} achieves the desired
decoupling of the emission phase space in unresolved limits. In the
collinear ($\eta \to 0$) and single-soft ($\xi_2 \to 0$) limits the
energy bound $E_{45,\mathrm{max}}$ simplifies to $\Emax=\frac{1}{2}
\beta^2 \mh$ and becomes independent of $\eta$ and $\xi_2$. In the
single-soft limit the momentum $q_5$ decouples from the energy-momentum
conserving $\delta$-function and the integrations over $\eta$ and
$\xi_2$ decouple from the resolved phase space. In the collinear limit
the momentum conservation depends only on the sum of momenta $q_{45}$,
which is the on-shell momentum of the massless parent parton of the
splitting and is independent of $\xi_2$. Again, the integrations over
$\eta$ and $\xi_2$ decouple from the resolved phase space. Note that the
collinear and the single-soft limits both yield the same resolved
configuration. Furthermore, in the double-soft limit, both gluons
decouple from the momentum conservation and the integrals over $\xi_1$,
$\xi_2$ and $\eta$ can be carried out for a fixed Born configuration.

\subsection{Pole vs. \texorpdfstring{\MSbar{}}{MSbar} Yukawa coupling}
\label{sec:pole_msbar_relation}
%%%
Already in the first calculation of the radiative corrections to the
Higgs boson decay rate to fermions discussed in
Ref.~\cite{Braaten:1980yq}, it has been recognised that the result
expressed in terms of the on-shell Yukawa coupling contains large
logarithms in the limit $\mb/\mh \ll 1$. It has also been shown there
that these large logarithms can be avoided by reexpressing the result in
terms of the \MSbar{} Yukawa coupling evaluated at the renormalisation
scale $\mu = \mh$.

The Yukawa coupling in the \MSbar{} scheme $\ybbar{\mu}$ is related to
the \MSbar{} mass $\mbbar{\mu}$ via
\begin{align}
  \ybbar{\mu} ={}&  (2\sqrt{2} G_F)^{1/2} \, \mbbar{\mu}
  \,,
\end{align}
where $\mu$ is the renormalisation scale.
Thus, the relation between the on-shell and \MSbar{} Yukawa couplings
can be deduced from the corresponding relation between the masses,
which we need up to
$\order{\as^2}$~\cite{Broadhurst:1991fy,Gray:1990yh}. It reads
\begin{align}
  \label{eq:yukawa-conversion}
  \yb^2
  ={}&
       \ybbarsq{\mu}
       \lb{
       1 + \asOnPi{\mu} r_1(\mb,\mu) + \asOnPi{\mu}^2 r_2(\mb,\mu)
       + \mathcal{O}(\as^3)
       }\rb
       \,,
\end{align}
where the coefficients $r_i(\mb,\mu)$ are presented in
\cref{sec:pole2msbar-coeff}.

%%%
Note that we only reexpress the overall Yukawa coupling in this way,
but we keep the mass dependence of the matrix elements and kinematical
invariants in terms of the pole mass, similar to
Ref.~\cite{Bernreuther:2018ynm}.

The total decay width and its expansion coefficients computed with the
\MSbar{} Yukawa coupling are denoted with a bar, i.e.
\begin{align}
  \label{eq:width-expansion-msbar}
  \overline{\Gamma}^{\bb{}}
  ={}&
       \overline{\Gamma}_{\LO}^{\bb{}}
       \lb{
       1
       + \asOnPi{} \overline{\gamma}_{1}^{\bb{}}
       + \asOnPi{}^2 \overline{\gamma}_{2}^{\bb{}}
       + \mathcal{O}(\as^3)
       }\rb\,,
\end{align}
where the expansion coefficients in the two schemes are related by
\begin{align}
  \label{eq:pole-msbar-relation-gamma}
  \overline{\gamma}_{1}^{\bb{}}
  ={}&
       \gamma_{1}^{\bb{}} + r_1
       \,,
  \\
  \overline{\gamma}_{2}^{\bb{}}
  ={}&
       \gamma_{2}^{\bb{}} + r_1 \gamma_{1}^{\bb{}} + r_2
       \,.
\end{align}

As discussed before, the large logarithmic corrections to the total
decay width can be mitigated by reexpressing the result in terms of the
running \MSbar{} mass of the $b$-quark.
However, in a fully differential calculation these logarithms partially
enter through corrections related to real emissions.
In this case, they arise during phase-space integration of the
emission. A priori, since the mass of the $b$-quark is small compared
to the Higgs mass, one could be worried about possible numerical
instabilities when working with a fully differential
calculation. However, it turns out that in our implementation they do
not pose serious numerical problems.

% -------------------------
\section{\texorpdfstring{\hbb{}}{H -> b bar} decay at NLO}
\label{sec:nlo}
In this section we briefly describe the calculation of the NLO QCD
corrections to the \hbb{} decay with massive $b$-quarks.
Although such a calculation is straightforward, we find it useful to
review it in order to clarify our notation and conventions. At this
order of perturbation theory we need to consider real~(R) and virtual
corrections~(V).

Nowadays the fully differential treatment of this decay mode can
easily be obtained using the
FKS~\cite{Frixione:1995ms,Frixione:1997np} or
Catani-Seymour~\cite{Catani:1996vz,Catani:1996jh,Catani:2002hc}
subtraction schemes. At NLO, our approach is essentially equivalent to
the FKS subtraction method.

\subsection{Real contribution}
\label{sec:nlo_real}
At NLO we consider one real emission in addition to the Born process,
which means that we need to integrate the function
\begin{align}
  \label{eq:FLM_bbg_nlo}
  \FLM{\bb{g}}
  ={}&
       \d\Phi_{\bb{g}}(q_1)
       \MEsq{0}{\bb{g}}
       \obs{\bb{g}}\,,
\end{align}
over the $\bb{g}$ phase space.  This integral is divergent in four
dimensions, due to the soft singularity of the gluon.
However, there are no collinear singularities since they are regulated
by the $b$-quark mass.

We define a projection operator that allows us to extract the soft
divergence and to regulate the limit. Given a quantity $A$ that depends
on the momenta, we define a projection operator for the soft limit of
momentum $q_4$ as
\begin{align}
  \label{eq:S4-def}
  \soft{4} A
  ={}&
       \lim_{\xi_1\to0} A
       \,,
\end{align}
where $\xi_1$ refers to the parametrisation of
\cref{eq:phsp_hbbg}.
We define the operator to act on all quantities to the right of the
$\soft{4}$ symbol, extracting the leading asymptotic behaviour in
$\xi_1$ of the quantity $A$ if the actual limit does not exist.

Denoting the identity operation by $I$, we can immediately write
\begin{align}
  \label{eq:nested_nlo}
  \FLM{\bb{g}}
  ={}&
     \FLM[(\ident-\soft{4})]{\bb{g}}
     +\FLM[\soft{4}]{\bb{g}}\,,
\end{align}
where the first term is now regularised in the soft limit and can be
integrated in four dimensions. The soft singularity is exposed in the
second term in \cref{eq:nested_nlo}, which therefore needs to be
evaluated in $d$ dimensions. Once the soft limit is taken, the only
remaining dependence on $\xi_1$ in $\FLM[\soft{4}]{\bb{g}}$ is the
leading behaviour $\xi_1^{-1-2\ep}$. Thus, the $\xi_1$ integration
becomes trivial and the soft singularity manifests itself as an
explicit $\ep^{-1}$ pole.

In order to show pole cancellation pointwise in the Born phase space,
we split the real emission contribution into two parts. The finite
contribution is given by
%%%texparser:start:contrib%%%
\begin{align}
  %%%texparser:LHS:dGam[R,F]%%%
  2\mh \, \dGamInt{\R}{\F}{\bb{g}}
  ={}&
  %%%texparser:stop%%%
       \FLMInt[(\ident-\soft{4})]{\bb{g}}
  \nonumber \\
  \label{eq:RF}
  ={}&
  %%%texparser:start:contrib%%%
       \la
       \FLM{\bb{g}}
       +\gs^2
       \CF
       \lp{
       \Sij{0}{22}{4}
       -2\Sij{0}{23}{4}
       +\Sij{0}{33}{4}
       }\rp
       \soft{4}
       \dq{4}
       \FLM{\bb{}}
       \ra
  \,.
\end{align}
%%%texparser:stop%%%
Here, we use the factorisation formula for the soft limit as discussed
in \cref{sec:fact-single-soft}, where also the eikonal factors
$\Sij{0}{ij}{k}$ are defined.
Moreover, we have the unresolved contribution, which contains the
integrated subtraction term and reads
%%%texparser:start:contrib%%%
\begin{align}
  %%%texparser:LHS:dGam[R,U]%%%
  2\mh \, \dGamInt{\R}{\U}{\bb{g}}
  ={}&
  %%%texparser:stop%%%
       \FLMInt[\soft{4}]{\bb{g}}
  \nonumber \\
  \label{eq:RU}
  ={}&
  %%%texparser:start:contrib%%%
       -
       \la
       \gs^2
       \CF
       \lp{
       \SijInt{0}{22}
       -2\SijInt{0}{23}
       +\SijInt{0}{33}
       }\rp
       \FLM{\bb{}}
       \ra
       \,,
\end{align}
%%%texparser:stop%%%
with the integrated eikonal factors $\SijInt{0}{ij}$ given in
\cref{sec:Sij0int}.
Here, the integral over the unresolved phase space of the gluon was
performed and we are only left with the phase-space integral over the
underlying Born process.

\subsection{Virtual contribution}
\label{sec:nlo_virt}
For the virtual contribution, the phase-space integration is the same
as that for the Born process, but we need to consider a one-loop
virtual amplitude.
Although this amplitude has an $\ep^{-1}$ pole, the singular part can
be written as a product of a tree-level matrix element and a
kinematics-dependent coefficient, as indicated in
\cref{sec:virtIR}. We have
\begin{align}
  2\Re\BraAmpl{0}{\bb{}}|\KetAmpl{1}{\bb{}}
  ={}&
       2\Re\BraAmpl{0}{\bb{}}|\KetFinRem{1}{\bb{}}
       +2\Re(\Zop{1}{\bb{}}) \BraAmpl{0}{\bb{}}|\KetAmpl{0}{\bb{}}
       \,,
\end{align}
where the term with the $\Zop{}{}$ operator contains an explicit
$\ep^{-1}$ pole, while the second term is finite.
As a shorthand we introduce
\begin{align}
  \label{eq:FLVfin-def}
  \FLVfin{\bb{}}
  ={}&
       \d\Phi_{\bb{}}(q_1)
       \asOnFourPi{}
       2\Re\BraAmpl{0}{\bb{}}|\KetFinRem{1}{\bb{}}
       \obs{\bb{}}
       \,.
\end{align}

Accordingly, we define two contributions to the virtual correction: the
virtual finite contribution
%%%texparser:start:contrib%%%
\begin{align}
  %%%texparser:LHS:dGam[V,F]%%%
  \label{eq:VF}
  2\mh \, \dGamInt{\V}{\F}{\bb{}}
  ={}&
       \FLVfinInt{\bb{}}
\end{align}
and the virtual unresolved contribution
\begin{align}
  %%%texparser:LHS:dGam[V,U]%%%
  \label{eq:VU}
  2\mh \, \dGamInt{\V}{\U}{\bb{}}
  ={}&
       \la
       \asOnFourPi{}
       2 \Re(\Zop{1}{\bb{}})
       \FLM{\bb{}}
       \ra
  \,.
\end{align}
%%%texparser:stop%%%
The expansion coefficients of the $\Zop{}{}$ operator are given in
\cref{eq:hbb_Zop1}.

\subsection{Pole cancellation}
\label{sec:nlo_pole_cancellation}
At this point we can combine all contributions that enter the NLO
calculation.
We remind the reader that the $\dGam{\R}{\F}{\bb{g}}$ and
$\dGam{\V}{\F}{\bb{}}$ terms are free of $\ep^{-1}$ poles, while the
$\dGam{\R}{\U}{\bb{g}}$ and $\dGam{\V}{\U}{\bb{}}$ terms feature
$\ep^{-1}$ poles.

Expanding the explicit results for the real unresolved contribution
from \cref{eq:RU} up to $\order{\ep^{-1}}$ we find
%%%texparser:start:poles%%%
\begin{align}
  %%%texparser:LHS:poles[R,U]%%%
  \label{eq:RU_explicit}
  2\mh \, \dGamInt{\R}{\U}{\bb{g}}
  ={}&
       \frac{1}{\ep}
       \lb{
       \asOnFourPi{}
       4\CF
       \lb{
       1
       +\frac{1+\beta^2}{2\beta}\log\lp{\frac{1-\beta}{1+\beta}}\rp
       }\rb
       \FLMInt{\bb{}}
       }\rb
       + \order{\ep^0}
  \,.
\end{align}
Analogously, the virtual unresolved contribution, defined in
\cref{eq:VU}, yields
\begin{align}
  %%%texparser:LHS:poles[V,U]%%%
  \label{eq:VU_explicit}
  2\mh \, \dGamInt{\V}{\U}{\bb{}}
  ={}&
       -\frac{1}{ \ep }
       \lb{
       \asOnFourPi{}
       4\CF
       \lb{
       1
       +\frac{1+\beta^2}{2\beta}\log\lp{\frac{1-\beta}{1+\beta}}\rp
       }\rb
       \FLMInt{\bb{}}
       }\rb
       \,.
\end{align}
%%%texparser:stop%%%
Note that the constant term in $\ep$ is absent since our definition of
the $\Zop{}{}$ operator of \cref{eq:Zop-def} includes only $\ep^{-1}$
poles.

Obviously, the poles of the two contributions cancel.
A crucial part of the argument is to notice that the
$\dGam{\R}{\U}{\bb{g}}$ part, after integrating out the real emission,
is a function of a Born-like phase-space configuration, as is the
$\dGam{\V}{\U}{\bb{}}$ contribution. This allows us to demonstrate the
pole cancellation for any point of the $\d\Phi_{\bb{}}$ phase space and
without specifying the explicit form of the \LO{} matrix element.

%-------------------------
\section{\texorpdfstring{\hbb{}}{H -> b bbar} decay at NNLO}
\label{sec:nnlo}

We now consider the NNLO QCD corrections to the \hbb{} process keeping
the full dependence on the $b$-quark mass.
%% NNLO h->bb contributions
To this end, we need to consider several contributions including
\begin{itemize}
\item the double-real contribution (RR) -- where the leading-order
  decay is accompanied by an emission of a pair of massless partons
  (\hbb{gg} and \hbb{\qq{}}) or an additional pair of $b$-quarks
  (\hbb{\bb{}});
\item the real-virtual contribution (RV) -- where we consider one-loop
  virtual corrections to the process \hbb{g};
\item the double-virtual contribution (VV) -- where we consider
  two-loop virtual corrections to the Born process \hbb{}.
\end{itemize}
Except for the \hbb{\bb{}} subprocess, all these contributions are
divergent in four dimensions, due to soft and collinear singularities.
For that reason, we follow the general method recapitulated in
\cref{sec:philosophy} and adopt a subtraction scheme that allows
us to regulate all singular limits and treat the divergent integrals
analytically in $d=4-2\ep$ dimensions.

%% h->bbbb contributions
Apart from these divergent contributions we distinguish the subprocess
\hbb{\bb{}} which enters the calculation at $\order{\as^2}$. Indeed,
this subprocess is finite in four dimensions because of the non-zero
$b$-quark mass. Hence, it does not require any regularisation. It is
calculated by directly integrating the squared tree-level amplitude over
the phase space $\d\Phi_{\bbbb}$. In our implementation we use the
sequential algorithm~\cite{Tanabashi:2018oca} to generate kinematic
configurations and the phase-space measure.

%% yb*yt contributions
Finally, a distinct class of corrections to the \hbb{} decay that
appears at second order of perturbation theory is related to Feynman
diagrams where the \hbb{} or \hbb{g} transition is induced by the
Higgs boson coupling to gluons via a top-quark loop.
This contribution is finite on its own and, hence, can be studied
separately -- we defer the discussion of these top-quark mediated
corrections to \cref{sec:ybyt}.

%% amplitudes
All tree-level amplitudes that we use in this paper are calculated
using the spinor-helicity formalism.\footnote{For a review, see e.g.
  Ref.~\cite{Dixon:1996wi}.} The treatment of massive external
particles follows along the lines of Appendix~A of
Ref.~\cite{Brucherseifer:2013iv}.
The one-loop amplitudes are calculated using a combination of
Passarino-Veltman reduction~\cite{Passarino:1978jh}, to express them
through one-loop scalar integrals, and spinor-helicity techniques, to
treat spinor structures appearing in the amplitudes.  The one-loop
scalar integrals are evaluated using the library
\QCDLoop{}~\cite{Ellis:2007qk,Carrazza:2016gav}. 
We assemble the two-loop using the two-loop scalar heavy-quark form
factor from Ref.~\cite{Ablinger:2017hst}; equivalent results can be
obtained using the expressions from Ref.~\cite{Bernreuther:2005gw}. The
form factor is expressed in terms of harmonic
polylogarithms~\cite{Remiddi:1999ew}, which we evaluate using
\HPLOG{}~\cite{Gehrmann:2001pz}.

%% black box cancellation
It is useful to stress that we show cancellation of all $\ep^{-1}$ poles
without referring to the specific form of the matrix elements. The
expressions for all amplitudes are only needed to calculate finite
corrections to the considered process and, hence, we restrict ourselves
to the construction of only four-dimensional matrix elements.

\subsection{Double-real contribution}
\label{sec:nnlo_RR}

In this section we only focus on the \hbb{gg} and \hbb{\qq{}}
subprocesses. The \hbb{\bb{}} process is completely finite on its own
and does not require any regularisation procedure. For the record we
write
\begin{align}
  \label{eq:nnlo_4b}
  2\mh \, \dGamInt{\bbbb}{}{}
  ={}&
       \FLMInt{ \bbbb }
       \,,
\end{align}
where the shorthand $\FLM{ \bbbb }$ is introduced in
\cref{eq:FLM_def}.

%% singular regions
Thanks to the non-zero $b$-quark mass, the singularity structure of
the double-real contribution is simple. Indeed, we only need to take
into account three possible limits:
\begin{itemize}
\item the soft limit (\soft{5}) -- where the energy of one of the partons
  vanishes, i.e. $\xi_2\to0$;
\item the double-soft limit(\dsoft{45}) -- where the energies of both
  additional partons vanish at a similar rate, i.e. $\xi_1\to0$;
\item the collinear limit (\coll{45}) -- where the momenta of the two
  additional partons become collinear to each other, i.e. $\eta\to0$.
\end{itemize}
The variables $\xi_1$, $\xi_2$ and $\eta$ refer to the parametrisation
introduced in \cref{eq:phsp_hbbgg}.

We define projection operators that allow us to extract divergences in
each of the singular regions. Given a quantity $A$ which depends on the
momenta of the $b$-quarks and gluons, we define the action of the
projection operators as follows
\begin{align}
  \label{eq:nnlo_projections}
  \dsoft{45} A ={}& \lim_{\xi_1\to 0} A\,,&
  \soft{5} A ={}& \lim_{\xi_2\to 0} A\,,&
  \coll{45}A ={}& \lim_{\eta\to 0} A\,.&
\end{align}
Again, we note that taking limits in \cref{eq:nnlo_projections}
should be understood as extracting the most singular part of the
quantity $A$ in a particular limit whenever the limit in the
conventional sense does not exist.

With these operators, we construct a nested subtraction formula which
extracts all singularities of the double-real contribution,
\begin{align}
  \FLM{\bb{gg}}
  ={}&
       \FLM[(\ident-\dsoft{45})]{\bb{gg}}
       +\FLM[\dsoft{45}]{\bb{gg}}
       \nonumber \\
  ={}&
       \FLM[(\ident-\soft{5})(\ident-\dsoft{45})]{\bb{gg}}
       +\FLM[\soft{5}(\ident-\dsoft{45})]{\bb{gg}}
       + \FLM[\dsoft{45}]{\bb{gg}}
       \nonumber \\
  \label{eq:nested_subtraction}
  ={}&
       \FLM[(\ident-\soft{5})(\ident-\dsoft{45})(\ident-\coll{45})]{\bb{gg}}
       +\FLM[\soft{5}(\ident-\dsoft{45})(\ident-\coll{45})]{\bb{gg}}
       \nonumber\\
     &
       +\FLM[(\ident-\soft{5})(\ident-\dsoft{45})\coll{45}]{\bb{gg}}
       +\FLM[\soft{5}(\ident-\dsoft{45})\coll{45}]{\bb{gg}}
       \nonumber\\
     &
       +\FLM[\dsoft{45}]{\bb{gg}}\,.
\end{align}
To derive \cref{eq:nested_subtraction}, we start by regularising the
double-soft singularity (\dsoft{45}), followed by further
regularisation of the single-soft limit (\soft{5}). In the last step
we introduce a subtraction term for the collinear singularity
(\coll{45}).
Moreover, as these operators commute with each other
\cite{Caola:2017dug}, we have a freedom to choose which limit to take
first. Note that in \cref{eq:nested_subtraction} we start with the
collinear limit, where appropriate, having in mind the simplicity of the
corresponding factorisation formulae.

We subdivide \cref{eq:nested_subtraction} into separate contributions
according to the final state multiplicity. This allows us to discuss
pole cancellation for each of these contributions separately.

%%% RRF: finite
The first term on the right-hand side of \cref{eq:nested_subtraction}
represents the fully regulated double-real contribution. Therefore, it
can be evaluated in four dimensions using standard numerical techniques.
We write
\begin{align}
  \label{eq:RRF_gg}
  2\mh \, \dGamInt{\RR}{\F}{ \bb{gg} }
  ={}&
       \FLMInt[(\ident-\soft{5})(\ident-\dsoft{45})(\ident-\coll{45})]{\bb{gg}}
       \,,
\end{align}
and similarly for the $\qq$ emission
\begin{align}
  \label{eq:RRF_qq}
  2\mh \, \dGamInt{\RR}{\F}{ \bb{\qq} }
  ={}&
       \nl \,
       \FLMInt[(\ident-\dsoft{45})(\ident-\coll{45})]{\bb{\qq}}
       +(q \leftrightarrow \bar{q})
       \,,
\end{align}
where $\nl$ is the number of massless quark flavours and the last term
corresponds to the phase-space region with $E_5 > E_4$. Note that in
case of the \qq{} emission we do not subtract the single-soft limit
(\soft{5}) since this limit is not singular in case of a $g\to\qq$
splitting. The explicit form of the subtraction terms generated by the
limit operators can be constructed using the factorisation formulae
collected in \cref{sec:fact-form}.

%%% RRSU: single-unresolved
The next three terms on the right-hand side of
\cref{eq:nested_subtraction} are regulated in the double-soft limit,
but they are evaluated in at least one of the other two limits
(\coll{45} or \soft{5}). As this leaves one of the real emissions
unresolved, we call this contribution single-unresolved.
Since we perform the integral over $\dq{5}$ analytically, we do not need
to further regulate the $\soft{5}$ and $\coll{45}$ limits. Thus,  we
rearrange these terms such that for the gluon emissions they read
%%%texparser:start:contrib%%%
\begin{align}
  %%%texparser:LHS:dGam[RR,SU,gg]%%%
  \label{eq:RRSU_gg}
  2\mh \, \dGam{\RR}{\SU}{ \bb{gg} }
  ={}&
  %%%texparser:stop%%%
       \bigl\langle
       \bigl(
       \soft{5}(\ident-\dsoft{45})(\ident-\coll{45})
       +(\ident-\soft{5})(\ident-\dsoft{45})\coll{45}
       \nonumber\\
     &
       +\soft{5}(\ident-\dsoft{45})\coll{45}
       \bigr)
       \FLM{\bb{gg}}
       \bigr\rangle
  \nonumber \\
  ={}&
       \left\langle
      \bigl(
      \soft{5}(\ident-\dsoft{45})
      +(\ident-\dsoft{45})\coll{45}
      -\soft{5}(\ident-\dsoft{45})\coll{45}
      \bigr)
       \FLM{\bb{gg}}
       \right\rangle
  \nonumber \\
  ={}&
  %%%texparser:start:contrib%%%
       \Bigl\langle 
       \gs^2
       \lp{
       \PijInt{0}{gg}
       -\PijSoftInt{0}{gg}
       +\CA( \SijInt{0}{24} + \SijInt{0}{34} - \SijInt{0}{23} )
       }\right.
       \nonumber\\
     &
       \left.{
       -\CF( \SijInt{0}{22} - 2\SijInt{0}{23} + \SijInt{0}{33} )
       }\rp
       (\xi_1/2)^{-2\ep}
       \FLM[(\ident-\soft{4})]{\bb{g}}
       \Bigr\rangle 
  \,.
\end{align}
For the \qq{} emission, we obtain
\begin{align}
  %%%texparser:LHS:dGam[RR,SU,qq]%%%
  \label{eq:RRSU_qq}
  2\mh \, \dGamInt{\RR}{\SU}{ \bb{\qq} }
    ={}&
  %%%texparser:stop%%%
       2 \nl \,
       \FLMInt[(\ident-\dsoft{45})\coll{45}]{\bb{\qq}}
  \nonumber \\
    ={}&
  %%%texparser:start:contrib%%%
         2 \nl
         \la
         \gs^2
       \PijInt{0}{\qq}
       (\xi_1/2)^{-2\ep}
         \FLM[(\ident-\soft{4})]{\bb{g}}
         \ra
  \,,
\end{align}
%%%texparser:stop%%%
where we use the fact that in the collinear limit any infrared-safe
observable must be symmetric under $q \leftrightarrow \bar{q}$, which
leads to the factor $2$ in \cref{eq:RRSU_qq}.
To get from the first to the second lines of
\cref{eq:RRSU_gg,eq:RRSU_qq}, we use the factorisation formulae given
in \cref{sec:fact-form} and integrate them over the unresolved phase
space $\dq{5}$, which yields the integrated splitting functions
$\PijInt{0}{ij}$ and the integrated eikonal factors $\SijInt{0}{ij}$.
Their calculation is described given in \cref{sec:subt_int}.  The
soft-regulated matrix element $\FLM[(\ident-\soft{4})]{\bb{g}}$ has
the same form as the real finite contribution at NLO, \cref{eq:RF}.

%%% RRDU: double-unresolved
Finally, the last term in \cref{eq:nested_subtraction} is evaluated
in the double-soft limit. In this case, both partons are unresolved and
we call this contribution double-unresolved. We find
%%%texparser:start:contrib%%%
\begin{align}
  %%%texparser:LHS:dGam[RR,DU,gg]%%%
  \label{eq:RRDU_gg}
  2\mh \, \dGamInt{\RR}{\DU}{ \bb{gg} }
    ={}&
  %%%texparser:stop%%%
      \FLMInt[\dsoft{45}]{\bb{gg}}
  \nonumber \\
  ={}&
  %%%texparser:start:contrib%%%
       \la
       \gs^4
       \DSoftInt{0}{gg}{q_2,q_3}
       \FLM{\bb{}}
       \ra
  \,,
  \\
  %%%texparser:LHS:dGam[RR,DU,qq]%%%
  \label{eq:RRDU_qq}
  2\mh \, \dGamInt{\RR}{\DU}{ \bb{\qq} }
  ={}&
  %%%texparser:stop%%%
       2 \nl \,
       \FLMInt[\dsoft{45}]{\bb{\qq}}
  \nonumber \\
    ={}&
  %%%texparser:start:contrib%%%
         2 \nl
         \la
         \gs^4
         \DSoftInt{0}{\qq}{q_2,q_3}
         \FLM{\bb{}}
         \ra
  \,.
\end{align}
%%%texparser:stop%%%
The double-soft emissions decouple from the hard process and the whole
term can be written as a product of the double-soft function and the
Born matrix element. After performing the integration over the momenta
of the soft partons, we denote the integrated double-soft function as
$\DSoftInt{0}{ij}{q_2,q_3}$. Details of this calculation are presented
in \cref{sec:SSij1int}.

\subsection{Real-virtual contribution}
\label{sec:nnlo_RV}

We need to consider the \hbb{g} amplitude at one-loop level and
integrate its product with the tree-level amplitude for the \hbb{g}
process over the phase space $\d\Phi_{\bb{g}}$.
In the following, we use the shorthand
\begin{align}
  \FLV{\bb{g}} &= 
    \d\Phi_{\bb{g}}(q_1)
    \asOnFourPi{}
    2\Re\BraAmpl{0}{\bb{g}}|\KetAmpl{1}{\bb{g}}
    \obs{\bb{g}}
    \,.
\end{align}

As for the real contribution at NLO, we regulate the soft singularity
using the soft limit operator $\soft{4}$ defined in \cref{eq:S4-def}
and find
\begin{align}
  \FLV{\bb{g}}
    &= \FLV[(\ident-\soft{4})]{\bb{g}}
       +\FLV[\soft{4}]{\bb{g}}
  \,.
\end{align}

We split the one-loop amplitude into a singular part, containing all
$\ep^{-1}$ poles from loop integrals, and a finite remainder, see
\cref{sec:virtIR}. We write
\begin{align}
  \label{eq:RV_ampl_split}
  2\Re\BraAmpl{0}{\bb{g}}|\KetAmpl{1}{\bb{g}}
    &=
       2\Re(\Zop{1}{\bb{g}})\BraAmpl{0}{\bb{g}}|\KetAmpl{0}{\bb{g}}
       +2\Re\BraAmpl{0}{\bb{g}}|\KetFinRem{1}{\bb{g}}
       \,,
\end{align}
where the coefficient of the $\Zop{}{}$ operator is given in
\cref{eq:hbbg_Zop1} and the second term contains the finite remainder of
the one-loop amplitude defined in \cref{eq:AmplFinRem_def}.

In total, we split the calculation into four contributions, which need
to be combined with the corresponding terms from the double-real and
double-virtual contributions to show pole cancellation.
The first one is given by the soft-regulated terms containing the
finite remainder of the \hbb{g} matrix element. It is free of
$\ep^{-1}$ poles and thus we refer to it as the real-virtual finite
contribution. It reads
%%%texparser:start:contrib%%%
\begin{align}
  %%%texparser:LHS:dGam[RV,F]%%%
  \label{eq:RVF}
  2\mh \, \dGamInt{\RV}{\F}{}
  ={}&
  %%%texparser:stop%%%
       \FLVfinInt[(\ident-\soft{4})]{\bb{g}}
  \nonumber \\
  ={}&
  %%%texparser:start:contrib%%%
       \Bigl\langle
       \FLVfin{\bb{g}}
       \nonumber\\
     &
       +
       \gs^2
       \CF( \Sij{0}{22}{4}-2\Sij{0}{23}{4}+\Sij{0}{33}{4} )
       \nonumber\\
     &
       \hphantom{-}
       \times
       \lp{
       \soft{4}\dq{4}\FLVfin{\bb{}}
       +
       \asOnFourPi{}
       \lb{\Rij{1}{23}{4}+Z_A^{(1)}+Z_{\as}^{(1)}}\rb_{\ep^0}
       \soft{4}\dq{4}\FLM{\bb{}}
       }\rp
       \Bigr\rangle
  \,,
\end{align}
%%%texparser:stop%%%
where the necessary formulae for the factorisation of the one-loop
matrix element in the soft limit are given in \cref{sec:fact-form} and
the the renormalisation constants $Z_A$ and $Z_{\as}$ are given in
\cref{sec:renorm}. The notation $[\ldots]_{\ep^0}$ indicates that the
$\order{\ep^0}$ term of the expression between the brackets should be
taken. We use $\FLVfin{\bb{g}}$ in analogy to the definition in
\cref{eq:FLVfin-def}.

The corresponding term containing the soft-regulated $\Zop{}{}$ operator
for the $\hbb{g}$ process is called real-virtual single-unresolved
contribution and is given by
%%%texparser:start:contrib%%%
\begin{align}
  %%%texparser:LHS:dGam[RV,SU]%%%
  \label{eq:RVSU}
  2\mh \, \dGamInt{\RV}{\SU}{}
  ={}&
  %%%texparser:stop%%%
       \la
       \asOnFourPi{}
       (\ident-\soft{4}) \,
       2\Re(\Zop{1}{\bb{g}}) \FLM{\bb{g}}
       \ra
  \nonumber \\
  ={}&
  %%%texparser:start:contrib%%%
       \Bigl\langle
       \asOnFourPi{}
       2\Re(\Zop{1}{\bb{g}})
       \FLM{\bb{g}}
       \nonumber\\
     &
       +\gs^2
       \asOnFourPi{}
       \CF
       ( \Sij{0}{22}{4}-2\Sij{0}{23}{4}+\Sij{0}{33}{4} )
       \nonumber\\
     &
       \hphantom{-}
       \times
       \lp{
       \lb{\Rij{1}{23}{4}+ Z_{\as}^{(1)} + Z_A^{(1)}}\rb_{\mathrm{poles}}
       +2\Re(\Zop{1}{\bb{}})
       }\rp
       \soft{4}\dq{4}\FLM{\bb{}}
       \Bigr\rangle
  \,,
\end{align}
%%%texparser:stop%%%
where the subscript ``poles'' refers to taking only the pole terms up to
$\order{\ep^{-1}}$ into account.

The remaining two contributions are evaluated in the soft limit and are
integrated over the unresolved phase space $\dq{4}$. The integrated
subtraction term containing the finite remainder carries the superscript
\FR{} and reads
%%%texparser:start:contrib%%%
\begin{align}
  %%%texparser:LHS:dGam[RV,FR]%%%
  \label{eq:RVFR}
  2\mh \, \dGamInt{\RV}{\FR}{}
  ={}&
  %%%texparser:stop%%%
       \FLVfinInt[\soft{4}]{\bb{g}}
  \nonumber \\
  ={}&
  %%%texparser:start:contrib%%%
       -
       \la
       \gs^2
       \CF
       \lp{ \SijInt{0}{22}-2\SijInt{0}{23}+\SijInt{0}{33} }\rp
       \FLVfin{\bb{}}
       \ra
  \,.
\end{align}
%%%texparser:stop%%%
Again, the $\SijInt{0}{ij}$ correspond to integrated eikonal factors
discussed in \cref{sec:Sij0int}.

Finally, the contribution containing the soft limit of the $\Zop{}{}$
operator integrated over the unresolved phase space is referred to as
the real-virtual double-unresolved contribution. We obtain
%%%texparser:start:contrib%%%
\begin{align}
  %%%texparser:LHS:dGam[RV,DU]%%%
  \label{eq:RVDU}
  2\mh \, \dGamInt{\RV}{\DU}{}
  ={}&
  %%%texparser:stop%%%
       \la
       \asOnFourPi{}
       \soft{4} \, 2\Re(\Zop{1}{\bb{g}}) \FLM{\bb{g}}
       \ra
  \nonumber \\
  ={}&
  %%%texparser:start:contrib%%%
       \Bigl\langle
       \gs^2
       \asOnFourPi{}
       \CF\lp{
       \RInt{1}
       -\lb{
       Z_{\as}^{(1)} + Z_A^{(1)}
       +2\Re(\Zop{1}{\bb{}})
       }\rb
       }\right.
       \nonumber
  \\ &
       \left.{
       \hphantom{\gs^2\asOnFourPi{}\CF}
       \times
       (\SijInt{0}{22}-2\SijInt{0}{23}+\SijInt{0}{33})
       }\rp
       \FLM{\bb{}}
       \Bigr\rangle
  \,.
\end{align}
%%%texparser:stop%%%
The integrated one-loop soft function $\RInt{1}$ is defined and
calculated in \cref{sec:Rij1int}.

\subsection{Double-virtual contribution}
\label{sec:nnlo_VV}

For the \hbb{} process, the virtual corrections are described by a
single form factor,
\begin{align}
  \label{eq:bb_FF}
  |\KetAmpl{}{\bb{}}
  &=
    F_s(\mh^2,\mb^2,\muR^2) |\KetAmpl{0}{\bb{}}
    \,.
\end{align}
In \cref{eq:bb_FF}, $|\KetAmpl{0}{\bb{}}$ is the tree-level \hbb{}
amplitude, and $F_s$ is the scalar heavy-quark form factor, which can
be computed perturbatively as
\begin{align}
  F_s(\mh^2,\mb^2,\muR^2)
  ={}&
       1
       +\asOnFourPi{\muR^2} F_s^{(1)}(\mh^2,\mb^2,\muR^2)
       \nonumber\\
     &
       \hphantom{1}
       +\asOnFourPi{\muR^2}^2 F_s^{(2)}(\mh^2,\mb^2,\muR^2)
       + \order{\as^3}
       \,.
\end{align}
The expansion coefficients $F_s^{(k)}$ depend on Higgs and $b$-quark
masses, the renormalisation scale $\muR$ and the regulator $\ep$. The
two-loop heavy-quark form factor was computed in
Ref.~\cite{Bernreuther:2005gw} up to $\ep^0$ terms and to higher
orders in $\ep$ in Refs.~\cite{Gluza:2009yy} and
\cite{Ablinger:2017hst}. In our calculation we use the results of
Ref.~\cite{Ablinger:2017hst}.

We again use the $\Zop{}{}$ operator to split the double-virtual
correction into divergent terms, that contain all $\ep^{-1}$ poles, and
a finite remainder.
In this case, given the functions $F_s^{(1)}(\mh^2,\mb^2,\muR^2)$ and
$F_s^{(2)}(\mh^2,\mb^2,\muR^2)$ we reconstruct the finite remainders
$|\KetFinRem{1}{\bb{}}$ and $|\KetFinRem{2}{\bb{}}$ using the
definition in \cref{eq:AmplFinRem_def}. We also checked that all
$\ep^{-1}$ poles of the form factor are in agreement with the
$\Zop{}{}$ operator prediction of \cref{eq:hbb_Zop1,eq:hbb_Zop2}.
The finite double-virtual contribution reads
%%%texparser:start:contrib%%%
\begin{align}
  %%%texparser:LHS:dGam[VV,F]%%%
  \label{eq:VVF}
  2\mh \, \dGamInt{\VV}{\F}{\bb{}}
  ={}&
       \Bigl\langle
       \asOnFourPi{}^2
       \phsp{\bb{}}
       \lb{
       2\Re\BraAmpl{0}{\bb{}}|\KetFinRem{2}{\bb{}}
       + \BraFinRem{1}{\bb{}}|\KetFinRem{1}{\bb{}}
       }\rb
       \obs{\bb{}}
       \Bigr\rangle
       \,.
\end{align}
%%%texparser:stop%%%

Furthermore, we distinguish two contributions that contain all
$\ep^{-1}$ poles. First, we have a term that is proportional to the
finite remainder of the one-loop \hbb{} matrix element. It is given by
%%%texparser:start:contrib%%%
\begin{align}
  %%%texparser:LHS:dGam[VV,FR]%%%
  \label{eq:VVFR}
  2\mh \, \dGamInt{\VV}{\FR}{}
  ={}&
       \Bigl\langle
       \asOnFourPi{}
       2\Re(\Zop{1}{\bb{}})
       \FLVfin{\bb{}}
       \Bigr\rangle
  \,.
\end{align}
%%%texparser:stop%%%
The other contribution containing $\ep^{-1}$ poles comes with the Born
\hbb{} matrix element; it reads
%%%texparser:start:contrib%%%
\begin{align}
  %%%texparser:LHS:dGam[VV,DU]%%%
  2\mh \, \dGamInt{\VV}{\DU}{}
  ={}&
       \Bigl\langle
       \asOnFourPi{}^2
       \lp{ 2\Re(\Zop{2}{\bb{}}) +{\Zop{1}{\bb{}}}^{\dagger}\Zop{1}{\bb{}} }\rp
       \FLM{\bb{}}
       \Bigr\rangle
  \,.
\end{align}
%%%texparser:stop%%%%
The expansion coefficients of the $\Zop{}{}$ operator are given in
\cref{eq:hbb_Zop1,eq:hbb_Zop2}.

\subsection{Pole cancellation}
\label{sec:nnlo_pole_cancellation}

After collecting formulae for all NNLO contributions in the preceeding
subsections, we demonstrate pole cancellation.
We begin by considering
$\dGam{}{\FR}{} = \dGam{\RV}{\FR}{} + \dGam{\VV}{\FR}{}$. From the
real-virtual contribution, \cref{eq:RVFR}, we obtain the pole term
%%%texparser:start:poles%%%
\begin{align}
  %%%texparser:LHS:poles[RV,FR]%%%
  \label{eq:RVFR_explicit}
  2\mh \, \dGamInt{\RV}{\FR}{\bb{g}}
  ={}&
       \frac{1}{\ep}
       \lb{
       \asOnFourPi{}
       4\CF
       \lb{
       1
       +\frac{1+\beta^2}{2\beta}\log\lp{\frac{1-\beta}{1+\beta}}\rp
       }\rb
       \FLVfinInt{\bb{}}
       }\rb
       + \order{\ep^0}
  \,,
\end{align}
and the explicit expansion of the double-virtual contribution,
\cref{eq:VVFR}, yields
\begin{align}
  %%%texparser:LHS:poles[VV,FR]%%%
  \label{eq:VVFR_explicit}
  2\mh \, \dGamInt{\VV}{\FR}{\bb{}}
  ={}&
       -\frac{1}{ \ep }
       \lb{
       \asOnFourPi{}
       4\CF
       \lb{
       1
       +\frac{1+\beta^2}{2\beta}\log\lp{\frac{1-\beta}{1+\beta}}\rp
       }\rb
       \FLVfinInt{\bb{}}
       }\rb
       \,,
\end{align}
%%%texparser:stop%%%
which obviously cancels in the sum with \cref{eq:RVFR_explicit}.

For the single-unresolved contribution we have
$\dGam{}{\SU}{} = \dGam{\RR}{\SU}{} + \dGam{\RV}{\SU}{}$. The
double-real contribution, \cref{eq:RRSU_gg,eq:RRSU_qq}, has an expansion
which can be cast into the form
%%%texparser:start:poles%%%
\begin{align}
  %%%texparser:LHS:poles[RR,SU]%%%
  \label{eq:RRSU_poles}
  \MoveEqLeft{2\mh \, \dGamInt{\RR}{\SU}{ \bb{gg} + \bb{\qq} }}
  \nonumber \\
  ={}&
       \Biggl\langle
       \asOnFourPi{}
       \Biggl[
       \frac{2 \CA}{\ep^2}
       +\frac{1}{\ep}
       \biggl[
       4\CF
       +\beta_0(\nl)
       +\frac{ \CA-2\CF }{ v_{23} } \log\lp{\frac{ 1+v_{23} }{ 1-v_{23}}}\rp
  \nonumber \\ &
       +2\CA\log\lp{\frac{ \mb \muR }{ 2\scprod{q_2}{q_4} }}\rp
       +2\CA\log\lp{\frac{ \mb \muR }{ 2\scprod{q_3}{q_4} }}\rp
       \biggr]
       \Biggr]
       \FLM[(\ident-\soft{4})]{\bb{g}}
       \Biggr\rangle
       +\order{\ep^0}
       \,.
\end{align}
A similar expansion holds for the real-virtual single-unresolved
contribution, \cref{eq:RVSU},
\begin{align}
  %%%texparser:LHS:poles[RV,SU]%%%
  \label{eq:RVSU_poles}
  \MoveEqLeft{2\mh \, \dGamInt{\RV}{\SU}{ \bb{g}}}
  \nonumber \\
  ={}&
       \Biggl\langle
       \asOnFourPi{}
       \Biggl[
       -\frac{2 \CA}{\ep^2}
       -\frac{1}{\ep}
       \biggl[
       4\CF
       +\beta_0(\nl)
       +\frac{ \CA-2\CF }{ v_{23} } \log\lp{\frac{ 1+v_{23} }{ 1-v_{23}}}\rp
       \nonumber
  \\ &
       +2\CA\log\lp{\frac{ \mb \muR }{ 2\scprod{q_2}{q_4} }}\rp
       +2\CA\log\lp{\frac{ \mb \muR }{ 2\scprod{q_3}{q_4} }}\rp
       \biggr]
       \Biggr]
       \FLM[(\ident-\soft{4})]{\bb{g}}
       \Biggr\rangle
       \,.
\end{align}
%%%texparser:stop%%%
As before, the poles of the two contributions cancel.

Finally, we turn to the double-unresolved contribution, given by
$\dGam{}{\DU}{} = \dGam{\RR}{\DU}{} + \dGam{\RV}{\DU}{} +
\dGam{\VV}{\DU}{}$. The real-virtual and double-virtual contributions
are known analytically, but we only have numerical results for the
double-real contribution at our disposal. Therefore, we demonstrate
pole cancellation numerically. We write the pole terms as
\begin{align}
  \dGamInt{}{\DU}{}
  ={}&
       \asOnFourPi{}^2
       \lb{
       \frac{ \partcoeff{}{\DU}{-3}}{\ep^3}
       +\frac{\partcoeff{}{\DU}{-2}}{\ep^2}
       +\frac{\partcoeff{}{\DU}{-1}}{\ep}
       }\rb
       \FLMInt{\bb{}}
       +\order{\ep^0}
  \,,
\end{align}
and further subdivide the pole coefficients according to colour factors,
\begin{align}
  \label{eq:DU_poles_colstr}
  \partcoeff{}{\DU}{k}
  ={}&
       \CF\CA \, \partcoeff{\CF\CA}{\DU}{k}
       +\CF^{\,2} \, \partcoeff{\CF^2}{\DU}{k}
       +\CF\TF\nl \, \partcoeff{\CF\TF\nl}{\DU}{k}
       +\CF\TF \, \partcoeff{\CF\TF}{\DU}{k}
       \,.
\end{align}
We list numerical values for these pole coefficients in
\cref{tab:nnlo_DU_poles}.
The pole cancellation reported in the last row occurs to at least 7 to
8 significant digits which proves pole cancellation in the
double-unresolved term.
%%%
\begin{table}[t]
\centering
\scalebox{0.90}{
\begin{tabular}{lrrrrrrrr}
  \toprule
  &
  $\partcoeff{\CF\CA}{\DU}{-3}$ &
  $\partcoeff{\CF\CA}{\DU}{-2}$ &
  $\partcoeff{\CF^2}{\DU}{-2}$ &
  $\partcoeff{\CF\TF\nl}{\DU}{-2}$ &
  $\partcoeff{\CF\CA}{\DU}{-1}$ &
  $\partcoeff{\CF^2}{\DU}{-1}$ &
  $\partcoeff{\CF\TF\nl}{\DU}{-1}$ &
  $\partcoeff{\CF\TF}{\DU}{-1}$ \\
  \midrule
  \RR{}  &
    $ -22.11 $ &
    $ -279.75 $ &
    $ +244.32 $ &
    $ +14.74 $ &
    $ -1777.55 $ &
    $ +2672.58 $ &
    $ +185.77 $ &
    $ 0 $ \\
  \RV{}  &
    $ +22.11 $ &
    $ +320.28 $ &
    $ -488.64 $ &
    $ -29.47 $ &
    $ +1732.44 $ &
    $ -2672.58 $ &
    $ -161.20 $ &
    $ +257.20 $ \\
  \VV{}  &
    $ 0 $ &
    $ -40.53 $ &
    $ +244.32 $ &
    $ +14.74 $ &
    $ +45.11 $ &
    $ 0 $ &
    $ -24.56 $ &
    $ -257.20 $ \\
  \midrule
  Sum &
    $ 10^{-13} $ &
    $ 10^{-10} $ &
    $ 10^{-8} $ &
    $ 10^{-11} $ &
    $ 10^{-6} $ &
    $ 10^{-6} $ &
    $ 10^{-5} $ &
    $ 0 $ \\
  Rel. canc. &
    $ 10^{-14} $ &
    $ 10^{-13} $ &
    $ 10^{-11} $ &
    $ 10^{-13} $ &
    $ 10^{-10} $ &
    $ 10^{-9} $ &
    $ 10^{-7} $ &
    $ 0 $ \\
  \bottomrule
\end{tabular}
}
\caption{Numerical values of the pole coefficients of the
  double-unresovled term as defined in \cref{eq:DU_poles_colstr}. The
  numerical values correspond to $\mb = 4.78~\gev$,
  $\mh = 125.09~\gev$ and the renormalisation scale is $\muR =
  3\mh$. Each column corresponds to a particular colour structure of a
  given $\ep$ pole. The three rows correspond to the double-real,
  real-virtual, and double-virtual contributions. In the last two
  rows, we report the absolute and relative level of cancellation
  after adding up $\RR+\RV+\VV$ contributions. The last row is
  normalised to the largest value of each column.
}
\label{tab:nnlo_DU_poles}
\end{table}

\subsection{Top-quark contribution to the \texorpdfstring{\hbb{}}{H -> b bar} decay}
\label{sec:ybyt}

\begin{figure}
  \centering
  \begin{subfigure}{0.45\textwidth}
    \centering
    \includegraphics[width=0.8\textwidth]{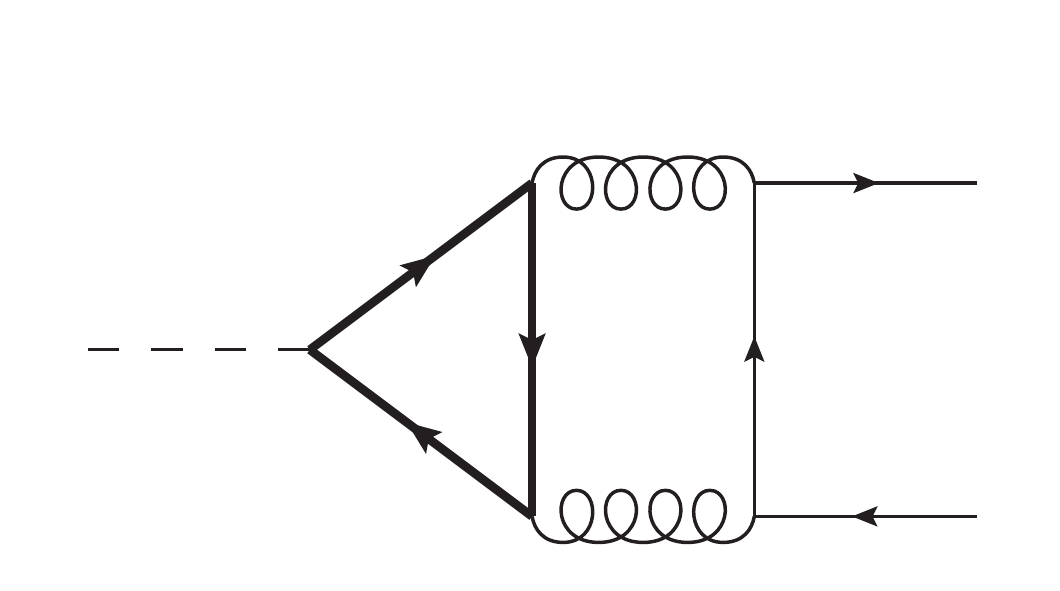}
    \caption{Two-loop $\hbb{}$ diagram}
    \label{fig:hbb2l_ybyt}
  \end{subfigure}
  \begin{subfigure}{0.45\textwidth}
    \centering
    \includegraphics[width=0.8\textwidth]{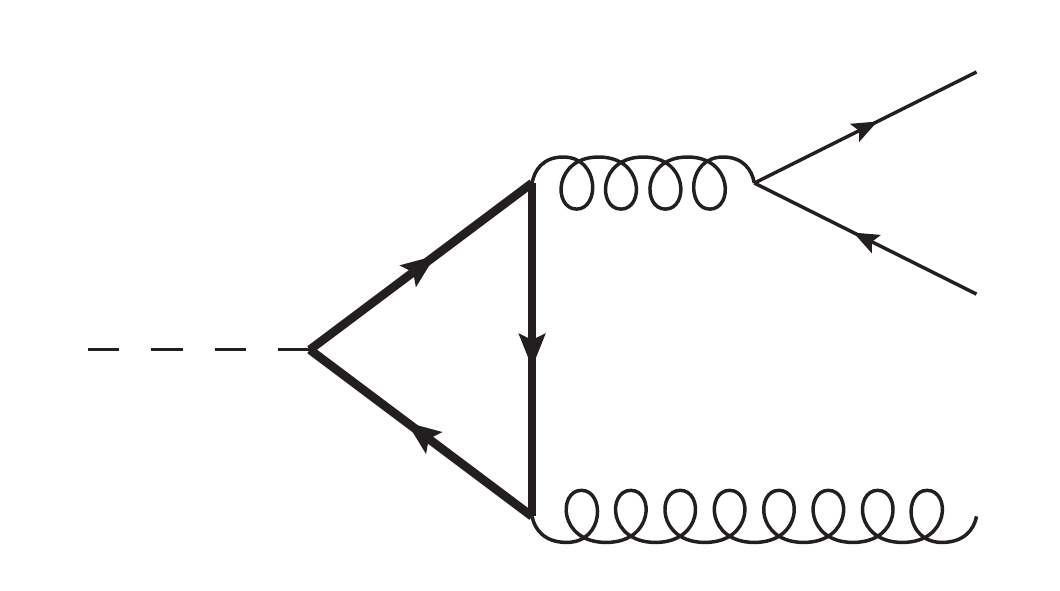}
    \caption{One-loop $\hbb{g}$ diagram}
    \label{fig:hbbg1l_ybyt}
  \end{subfigure}
  \caption{The top-Yukawa contributions to the \hbb{} and \hbb{g}
    amplitudes. The solid thick lines represent a top quark while the
    thin solid lines denote the external $b$-quarks.}
  \label{fig:ybyt-diags}
\end{figure}

An additional contribution that enters the \hbb{} decay process at
$\mathcal{O}(\as^2)$ is related to diagrams where the Higgs boson
couples to two gluons via a top-quark loop and the final state
$b$-quarks are generated by a gluon splitting. The corresponding
diagrams are shown in \cref{fig:hbb2l_ybyt,fig:hbbg1l_ybyt} and their
contributions to the decay rate are separately finite.
%%%
This contribution to the total \hbb{} width has been computed in
Refs.~\cite{Larin:1995sq,Chetyrkin:1995pd} as an expansion in powers of
$\xt = (\mh^2/\mt^2)$.
An exact result, incorporated into a fully differential calculation,
has been recently published in Ref.~\cite{Primo:2018zby}. At the level
of the total decay width, the differences between the approximate and
the exact results are very small for realistic Higgs and top-quark
masses. Since we aim at testing our calculation against results
reported in Ref.~\cite{Bernreuther:2018ynm}, we will use the result of
Ref.~\cite{Larin:1995sq} to include the top-quark contribution.

%%%
Note that this contribution should naively scale with one power of
$\yb$ and one power of $\yt$, hence we refer to this as the $\yb\yt$
contribution. However, the amplitudes of \cref{fig:ybyt-diags} can
interfere with the respective Born amplitudes only if a helicity flip
occurs on one of the $b$-quark lines. This mechanism provides an
additional power of $\mb$.
Moreover, the top-quark loop leads to a suppression factor of
$1/m_t$, which makes the overall scaling of the $\yb\yt$ contribution
similar to all other terms, i.e. proportional to a square of the
$b$-quark mass.

We write the $\yb\yt$ contribution to the total decay width
as~\cite{Larin:1995sq}
\begin{align}
  \label{eq:ybyt_corr}
  \Gamma^{\yb\yt}
  ={}&
       \Gamma_{\LO}
       \,
       \asOnPi{}^2
       \gamma_{2}^{\yb\yt}
       \,,
\end{align}
where
\begin{align}
  \label{eq:ybyt_LRV95}
  \gamma_{2}^{\yb\yt}
  ={}&
       \beta^{-3}
       \lp{
       f_{2}^{S,0}
       + \xt f_{2}^{S,1}
       + \xt^{\,2} f_{2}^{S,2}
       + \xt^{\,3} f_{2}^{S,3}
       + \mathcal{O}(\xt^{\,4})
       }\rp
\end{align}
and the coefficients $f_{2}^{S,k}$ are given in Eq.~(2) of
Ref.~\cite{Larin:1995sq}. The factor of $\beta^{-3}$ appears since the
normalisation factor in \cref{eq:ybyt_corr} involves the LO width for
the massive $b$-quarks, in contrast to Ref.~\cite{Larin:1995sq}.

Since the contribution in \cref{eq:ybyt_corr} starts only at
$\mathcal{O}(\as^2)$, changes due to decoupling,
\cref{eq:as-decoupling}, or due to translating the on-shell Yukawa
coupling to the \MSbar{} scheme, \cref{eq:yukawa-conversion}, are of
higher order in $\as$ and are discarded.

The real-virtual amplitude related to the diagram in
\cref{fig:hbbg1l_ybyt} is computed using the same techniques as the
remaining one-loop amplitudes. The double-virtual amplitude of
\cref{fig:hbb2l_ybyt} is obtained from the result of
Ref.~\cite{Larin:1995sq}, see \cref{eq:ybyt_corr}, by subtracting the
integrated real-virtual contribution of \cref{fig:hbbg1l_ybyt}.

% -------------------------
\section{Results}
\label{sec:results}
In this section, we summarise our calculation and present results for
the total decay width of the \hbb{} decay and a selection of jet rates.
We compare these predictions to results already available in the
literature.
The main goal is to scrutinise our calculation as much as possible to
ensure its correctness.
%

%%% setup
The value of the strong coupling is set to $\as(\mz)=0.1181$ with
$\mz=91.1876~\gev$~\cite{Tanabashi:2018oca}. The evolution is
performed at two-loop order with five active flavours using the
package \RunDec{}~\cite{Chetyrkin:2000yt,Herren:2017osy}.
We use a value for the Higgs boson mass of
$\mh = 125.09~\gev$~\cite{Aad:2015zhl}.
As a starting point for the $b$-quark mass, we use
$\mbbar{\mu=\mbbar{}} = 4.18~\gev$~\cite{Tanabashi:2018oca}. From this
we calculate the pole mass $\mb = 4.78~\gev$, using the two-loop
matching formula at a matching scale of $\mu=\mbbar{\mbbar{}}$
implemented in the package \RunDec{}.
For the value of the Fermi constant we use
$G_F = 1.166378 \times 10^{-5}~\gev^{-2}$~\cite{Tanabashi:2018oca}.
The central renormalisation scale is taken to be equal to the mass of
the Higgs boson, $\muR = \mh$, and for the scale uncertainty
estimation we vary it by a factor of $1/2$ and $2$.

\subsection{Overview of the calculation}
We combine all contributions discussed in \cref{sec:nlo,sec:nnlo} to
obtain the full NLO and NNLO results.
At leading order in QCD we have
\begin{align}
  2\mh \, \dGamInt{\LO}{\bb{}}{}
  ={}&
            \FLMInt{\bb{}}
       \,.
\end{align}
%%% NLO
At NLO we combine real and virtual corrections
\begin{align}
  \label{eq:result_NLO_Fkin}
  \dGamInt{\delta\NLO}{\bb{}}{}
  ={}&
       \dGamInt{\R}{}{}
       + \dGamInt{\V}{}{}
       \,,
\end{align}
which contain finite and unresolved parts as discussed in
\cref{sec:nlo}.
%%% NNLO
The situation is slightly more complicated at NNLO where there are
more ingredients entering the final result. The result reads
\begin{align}
  \label{eq:result_NNLO_Fkin}
  \dGamInt{\delta\NNLO}{\bb{}}{}
  ={}&
       \dGamInt{\VV}{}{}
       +\dGamInt{\RV}{}{}
       +\dGamInt{\RR}{}{}
       +\dGamInt{\bbbb}{}{}
       +\dGamInt{\bb{}}{\yb\yt}{}
       \,.
\end{align}
The double-virtual, real-virtual and double-real contributions consist
of finite, single- and double-unresolved parts as outlined in
\cref{sec:nnlo}.

\subsection{Total width of the \texorpdfstring{\hbb{}}{H -> b bar} decay at NNLO}
We start by considering the total width of the decay $\hbb{}$; this
implies using $\obs{\bb{X}}=1$.
We first present the results using the on-shell Yukawa coupling to
compare the NNLO decay width to a result in the large Higgs mass limit
from Ref.~\cite{Harlander:1997xa}.
Afterwards, we present the results expressed in terms of the \MSbar{}
Yukawa coupling.

\paragraph{Total width using the on-shell Yukawa coupling}
%%% 
We compare our results for the total decay width,
\cref{eq:width-expansion}, to the predictions of
Ref.~\cite{Harlander:1997xa}, which has been obtained as an expansion in
$\mb^2/\mh^2$. This result was derived for the scenario of a decay of a
heavy scalar boson into a pair of top quarks. Nevertheless, it can
immediately be translated into the \hbb{} decay rate if we neglect the
top-loop mediated contribution, discussed in \cref{sec:ybyt}.
Since the results of Ref.~\cite{Harlander:1997xa} are presented in
terms of the on-shell Yukawa coupling, we perform the comparison in
this scheme.

To scrutinise our results, we split the NLO and NNLO coefficients into
independent colour structures
\begin{align}
  \label{eq:width-expansion-colour-str}
  \gamma_{1}^{\bb{}}
  ={}&
       \CF \gamma_{1}^{\CF}
       \,,
       \\
  \gamma_{2}^{\bb{}}
  ={}&
        \CF^2     \gamma_{2}^{\CF^2}
       +\CF\CA    \gamma_{2}^{\CF\CA}
       +\CF\TF\nl \gamma_{2}^{\CF\TF\nl}
       +\CF\TF    \gamma_{2}^{\CF\TF}
       \,.
\end{align}
Our findings are summarised in
\cref{tab:total_width_colstr_HS_comparison}.
The first row lists predictions of Ref.~\cite{Harlander:1997xa} which
are compared to our predictions in the second row.
We see a remarkable consistency between the two; the discrepancies are
at most at the level of our numerical errors. This good agreement is
related to the smallness of the expansion parameter in case of a Higgs
decay to $b$-quarks, $(\mb^2/\mh^2) \approx 0.00146$.

%%%
\begin{table}
\centering
\scalebox{1.0}{
\begin{tabular}{llllll}
  \toprule
  &
  $\gamma_{1}^{\CF}$ &
  $\gamma_{2}^{\CF^2}$ &
  $\gamma_{2}^{\CF\CA}$ &
  $\gamma_{2}^{\CF\TF\nl}$ &
  $\gamma_{2}^{\CF\TF}$ \\
  \midrule
  Ref.~\cite{Harlander:1997xa} &
  $-7.446648$ &
  $+19.4192$ &
  $-53.5558$ &
  $+18.6286$ &
  $+14.7946$ \\
  Our res. &
  $-7.446648(7)$ &
  $+19.4199(10)$ &
  $-53.5557(20)$ &
  $+18.6283(2)$ &
  $+14.7945(1)$ \\
  \bottomrule
\end{tabular}
}
\caption{The results for the NLO and NNLO coefficients of the total
  decay width split into independent colour structures.
  The renormalisation scale is set to $\muR=\mh$.
  The uncertainties quoted for our results correspond to errors
  from numerical integration.
  We note that the results do not include the $\yb\yt$ contribution.
}
\label{tab:total_width_colstr_HS_comparison}
\end{table}
Note that the numerical values of the NNLO coefficients are much
larger than those of the NLO coefficients. Since this is only
partially compensated by the additional power of the strong coupling,
$(\as/\pi)$, the NNLO correction in this scheme still amounts to a
sizeable change of the total decay width.
This behaviour is closely related to large quasi-collinear logarithms,
$L = \log(\mb/\mh)$, discussed in \cref{sec:pole_msbar_relation}.
As already indicated, this issue can be mitigated by reexpressing
the results in terms of the \MSbar{} Yukawa coupling.

\paragraph{Total width using the \MSbar{} Yukawa coupling}
\label{sec:tot-width-msbar}
We now express the total decay width in terms of the \MSbar{}
coupling, as defined in \cref{eq:width-expansion-msbar}.
We use $\mbbar{\mu=\mbbar{}} = 4.18~\gev$ as an input
parameter~\cite{Tanabashi:2018oca} and evolve the \MSbar{} mass with
the \RunDec{} package~\cite{Chetyrkin:2000yt,Herren:2017osy}. We
obtain
\begin{align}
  \mbbar{ \tfrac{1}{2}\mh } ={}& 2.9814~\gev\,, &
  \mbbar{             \mh } ={}& 2.8095~\gev\,, &
  \mbbar{            2\mh } ={}& 2.6641~\gev\,. &
\end{align}
The evolution is performed with $\nf=5$ active flavours at two-loop
order.
We use these values to evaluate the \MSbar{} Yukawa coupling.
The NLO and NNLO coefficients together with a prediction for a total
decay width are presented in \cref{tab:total_width_msbar}.
We first discuss the results without the top-quark contribution,
described in \cref{sec:ybyt}.
For comparison, we also include the NLO and NNLO coefficients obtained
analytically in the limit of massless $b$-quarks given in
Ref.~\cite{Chetyrkin:1996sr}.
%%%
\begin{table}[tb]
\centering
\scalebox{1.0}{
\begin{tabular}{llll}
  \toprule
  $\muR$&
  $\tfrac{1}{2}\mh$ &
  $\mh$ &
  $2\mh$ \\
  \midrule
  $\overline{\gamma}_{1}^{\bb{}}$ (our res.) &
  $+3.023597(10)$ &
  $+5.796203(15)$ &
  $+8.568783(11)$ \\
  $\overline{\gamma}_{1}^{\bb{}}$ (Ref.~\cite{Bernreuther:2018ynm}) &
  $+3.024$ &
  $+5.798$ &
  $+8.569$ \\
  $\overline{\gamma}_{1}^{\bb{}}$ (Ref.~\cite{Chetyrkin:1996sr}, $\mb=0$) &
  $+2.8941$ &
  $+5.6667$ &
  $+8.4393$ \\
  \midrule
  $\overline{\gamma}_{2}^{\bb{}}$ (our res., w/o $\yb\yt$)&
  $-3.2466(31)$ &
  $+30.4376(33)$ &
  $+79.1755(38)$ \\
  $\overline{\gamma}_{2}^{\bb{}}$ (our res., with $\yb\yt$)&
  $ +3.7123(31)$ &
  $+37.3965(33)$ &
  $+86.1345(38)$ \\
  $\overline{\gamma}_{2}^{\bb{}}$ (Ref.~\cite{Bernreuther:2018ynm}, with $\yb\yt$) &
  $+3.685$ &
  $+37.371$ &
  $+86.112$ \\
  $\overline{\gamma}_{2}^{\bb{}}$ (Ref.~\cite{Chetyrkin:1996sr}, $\mb=0$) &
  $-3.8368$ &
  $+29.1467$ &
  $+77.1844$ \\
  \midrule
  $\overline{\Gamma}_{\LO}^{\bb{}}~[\mev]$ &
  $+2.17005$ &
  $+1.92702$ &
  $+1.73274$ \\
  %\hline
  $\overline{\Gamma}_{\NLO}^{\bb{}}~[\mev]$ &
  $+2.43161$ &
  $+2.32781$ &
  $+2.21731$ \\
  %\hline
  $\overline{\Gamma}_{\NNLO}^{\bb{}}~[\mev]$ (w/o $\yb\yt$)&
  $   +2.42041(1)$ &
  $   +2.40333(1)$ &
  $   +2.36344(1)$ \\
  $\overline{\Gamma}_{\NNLO}^{\bb{}}~[\mev]$ (with $\yb\yt$)&
  $   +2.44441(1)$ &
  $   +2.42059(1)$ &
  $   +2.37628(1)$ \\
  \bottomrule
\end{tabular}
}
\caption{
  The results for the LO, NLO and NNLO total decay width.
  The total width is calculated using our results for the expansion coefficients,
  $\overline{\gamma}_{1}^{\bb{}}$ and $\overline{\gamma}_{2}^{\bb{}}$.
  For comparison we include corresponding results from
  Ref.~\cite{Bernreuther:2018ynm}. We also provide results in the limit of
  massless $b$-quarks from Ref.~\cite{Chetyrkin:1996sr}, which do not
  contain the $\yb\yt$ contribution.
  The uncertainties quoted for our results correspond to errors from
  numerical integration.  }
\label{tab:total_width_msbar}
\end{table}

We see a reasonably good perturbative convergence of the predictions
for the total decay width, $\overline{\Gamma}^{\bb{}}$.
At the central scale, the NNLO corrections change the NLO result by a
few percent and the NNLO prediction stays within the NLO scale
uncertainties.
Since the large quasi-collinear logarithms are removed by switching to
the \MSbar{} Yukawa coupling, the mass corrections are not large. The
difference between the NNLO coefficient of the massless and massive
predictions is at the level of about $4\%$ at the central scale,
$\muR=\mh$.

Nevertheless, the fully massive treatment of $b$-quarks is desirable
because the corrections related to the top-quark Yukawa coupling
cannot be incorporated into a fully differential calculation with
massless $b$-quarks~\cite{Caola:2017xuq}. Furthermore, it is also
important to study the impact of mass effects on the kinematical
distributions related to \hbb{} decay.

%%%%%%%%%%%%%%%%%%%%%%%%%%%%%%%%%%%%%%%%%%%%%%%%%%%%%%%%%%%%%%%%%%%%%%%%
\paragraph{Total decay width including the top-quark contribution}
Finally, we incorporate the top-quark contribution into our
predictions, according to the discussion presented in
\cref{sec:ybyt}. We set the top-quark pole mass to
$\mt = 173.34~\gev$~\cite{Bernreuther:2018ynm}.
The NNLO coefficient corresponding to top-quark induced contribution,
\cref{eq:ybyt_LRV95}, is then $\gamma_{2}^{\yb\yt} = 6.95895$ and is
independent of the renormalisation scale.
Our predictions are included in \cref{tab:total_width_msbar}.

We see that inclusion of the $\yb\yt$ contribution increases the total
width by about $1.7\%$ at the central renormalisation scale. A major
part of this correction, about $85\%$, comes from the real-virtual
diagrams of the $\yb\yt$ contribution, see \cref{fig:hbbg1l_ybyt}.

Comparing our results for the NLO and NNLO coefficients of the total
width to those given in Ref.~\cite{Bernreuther:2018ynm}, we find
excellent agreement at NLO. The NNLO coefficients agree at the level
of at least $0.7\%$ for $\muR=\mh/2$ and at the sub-permill level for
the other scales. Note that for
$\overline{\gamma}_2^{\bb{}}(\muR=\mh/2)$ there are large
cancellations when converting from the on-shell to the \MSbar{}-scheme
Yukawa coupling.

%%%%%%%%%%%%%%%%%%%%%%%%%%%%%%%%%%%%%%%%%%%%%%%%%%%%%%%%%%%%%%%%%%%%%%%%
\subsection{Jet rates for \texorpdfstring{\hbb{}}{H -> b bar} at NNLO}
\label{sec:jet-rates-results}
We now turn to a discussion of jet rates which provides another
stress test of our calculation.
We employ the Durham jet algorithm~\cite{Catani:1991hj} with the
default recombination scheme of the parton momenta,
$k_{(ij)} = k_i + k_j$. We use the
{\texttt{FastJet}}~\cite{Cacciari:2011ma} implementation. We consider
two cases of the clustering sequence with $\ycut=0.01$ and
$\ycut=0.05$, to facilitate a numerical comparison against
Ref.~\cite{Bernreuther:2018ynm}.

Similar to \cref{eq:width-expansion-msbar}, we define an expansion of
the differential quantities as
\begin{align}
  \label{eq:jet-rates-expansion-msbar}
  \overline{\Gamma}^{\bb{}}(\mathrm{obs})
  ={}&
       \overline{\Gamma}_{\LO}^{\bb{}}
       \,
       \lb{
       \overline{\gamma}_{0}^{\bb{}}(\mathrm{obs})
       + \asOnPi{} \overline{\gamma}_{1}^{\bb{}}(\mathrm{obs})
       + \asOnPi{}^2 \overline{\gamma}_{2}^{\bb{}}(\mathrm{obs})
       + \mathcal{O}(\as^3)
       }\rb\,,
\end{align}
where ``$\mathrm{obs}$'' denotes a generic observable and the \MSbar{}
quantities are related to their counterparts in the pole scheme via
\begin{align}
  \overline{\gamma}_{0}^{\bb{}}(\mathrm{obs})
  ={}&
       \gamma_{0}^{\bb{}}(\mathrm{obs})
       \,,
  \\
  \overline{\gamma}_{1}^{\bb{}}(\mathrm{obs})
  ={}&
       \gamma_{1}^{\bb{}}(\mathrm{obs}) + r_1 \gamma_{0}^{\bb{}}(\mathrm{obs})
       \,,
  \\
  \overline{\gamma}_{2}^{\bb{}}(\mathrm{obs})
  ={}&
       \gamma_{2}^{\bb{}}(\mathrm{obs}) + r_1 \gamma_{1}^{\bb{}}(\mathrm{obs}) + r_2 \gamma_{0}^{\bb{}}(\mathrm{obs})
       \,.
\end{align}
Note that the sum of expansion coefficients of all relevant jet
multiplicities, $\overline{\gamma}_{i}^{\bb{}}(n\mathrm{jet})$, yields
the expansion coefficient of the total decay width
$\overline{\gamma}_{i}^{\bb{}}$.

%%% jet-rates results
We keep using the setup presented in \cref{sec:tot-width-msbar} and
report results for the two-, three- and four-jet rates in
\cref{tab:jet-rates-msbar}.
We present these results with and without top-quark contributions
whenever relevant, i.e. for the two- and three-jet rate NNLO
coefficients.
%%% 
\begin{table}[tb]
  \centering
    \scalebox{0.85}{
      \begin{tabular}{lllllll}
        \toprule
        &
          \multicolumn{3}{c}{$\ycut=0.01$}
        &
          \multicolumn{3}{c}{$\ycut=0.05$}
        \\
        %\midrule
        \cmidrule(r{0.2em}){2-4} \cmidrule(l{0.2em}){5-7}
        $ \muR $&
        $ \tfrac{1}{2}\mh$ &
        $ \mh$ &
        $ 2\mh$ &
        $ \tfrac{1}{2}\mh$ &
        $ \mh$ &
        $ 2\mh$ \\
        \midrule
        $ \overline{\gamma}_{1}^{\bb{}}(\mathrm{2jet})$&
        $ -5.0559(1)$&
        $ -2.2832(1)$&
        $ +0.4894(1)$&
        $ +0.2903(1)$&
        $ +3.0629(1)$&
        $ +5.8355(1)$\\
        $ \overline{\gamma}_{1}^{\bb{}}(\mathrm{2jet})$ (Ref.~\cite{Bernreuther:2018ynm})&
        $ -5.055$&
        $ -2.282$&
        $ +0.490$&
        $ +0.291$&
        $ +3.063$&
        $ +5.836$\\
        \midrule
       $ \overline{\gamma}_{2}^{\bb{}}(\mathrm{2jet})$ (w/o $\yb\yt$)&
        $ -60.50(3)$&
        $ -70.68(1)$&
        $ -65.83(2)$&
        $ -25.42(1)$&
        $  -6.59(1)$&
        $ +27.31(1)$\\
        $ \overline{\gamma}_{2}^{\bb{}}(\mathrm{2jet})$ (with $\yb\yt$)&
        $ -56.40(3)$&
        $ -66.58(1)$&
        $ -61.73(2)$&
        $ -19.52(1)$&
        $  -0.69(1)$&
        $ +33.21(1)$\\
        $ \overline{\gamma}_{2}^{\bb{}}(\mathrm{2jet})$ (Ref.~\cite{Bernreuther:2018ynm})&
        $ -56.351$&
        $ -66.532$&
        $ -61.658$&
        $ -19.496$&
        $  -0.650$&
        $ +33.250$\\
        \midrule
        $\overline{\gamma}_{1}^{\bb{}}(\mathrm{3jet})$&
        $ +8.0794(1)$&
        $ +8.0794(1)$&
        $ +8.0794(1)$&
        $ +2.7333(1)$&
        $ +2.7333(1)$&
        $ +2.7333(1)$\\
        $\overline{\gamma}_{1}^{\bb{}}(\mathrm{3jet})$ (Ref.~\cite{Bernreuther:2018ynm})&
        $ +8.079$&
        $ +8.079$&
        $ +8.079$&
        $ +2.733$&
        $ +2.733$&
        $ +2.733$\\
        \midrule
        $ \overline{\gamma}_{2}^{\bb{}}(\mathrm{3jet})$ (w/o $\yb\yt$)&
        $  +34.09(3)$&
        $  +77.96(1)$&
        $ +121.84(2)$&
        $  +21.25(1)$&
        $  +36.09(1)$&
        $  +50.94(1)$\\
        $ \overline{\gamma}_{2}^{\bb{}}(\mathrm{3jet})$ (with $\yb\yt$)&
        $  +36.95(3)$&
        $  +80.82(1)$&
        $ +124.70(2)$&
        $  +22.30(1)$&
        $  +37.15(1)$&
        $  +51.99(1)$\\
        $ \overline{\gamma}_{2}^{\bb{}}(\mathrm{3jet})$ (Ref.~\cite{Bernreuther:2018ynm})&
        $  +36.873$&
        $  +80.741$&
        $ +124.609$&
        $  +22.256$&
        $  +37.096$&
        $  +51.937$\\
        \midrule
        $\overline{\gamma}_{2}^{\bb{}}(\mathrm{4jet})$&
        $ +23.164(1)$&
        $ +23.163(1)$&
        $ +23.163(1)$&
        $ +0.9323(1)$&
        $ +0.9322(1)$&
        $ +0.9322(1)$\\
        $\overline{\gamma}_{2}^{\bb{}}(\mathrm{4jet})$ (Ref.~\cite{Bernreuther:2018ynm})&
        $ +23.163$&
        $ +23.163$&
        $ +23.163$&
        $ +0.926$&
        $ +0.926$&
        $ +0.926$\\
        \bottomrule
      \end{tabular}
    }
  \caption{The jet rates expansion coefficients
    $\overline{\gamma}_{i}^{\bb{}}(\mathrm{obs})$ as defined in
    \cref{eq:jet-rates-expansion-msbar} and computed using the
    Durham clustering algorithm with $\ycut = 0.01$ and $\ycut = 0.05$
    for various choices of the renormalisation scale. Whenever
    necessary, we report results without and with top-quark
    contributions.
    }
  \label{tab:jet-rates-msbar}
\end{table}
%%%
We see that the NLO coefficient of the three-jet rate and the NNLO
coefficient of the four-jet rate are scale-independent since they
involve only tree-level subprocesses.
We also observe a migration of events from higher to lower jet
multiplicities when jet-cut parameter is increased.

%%%
For the NLO and NNLO coefficients of the jet rates we see agreement
between our result and that of Ref.~\cite{Bernreuther:2018ynm} at the
level of $0.1\%$ to $0.2\%$ for all quantities considered for
$\ycut = 0.01$. Similar agreement is observed for $\ycut = 0.05$ with
the exception of
$\overline{\gamma}_2^{\bb{}}(2\mathrm{jet},\muR=\mh)$, which however
suffers from large cancellations in the course of the on-shell to
\MSbar{}-scheme conversion, and the four-jet rates.

%%% 
\begin{table}
  \centering
  \begin{tabular}{lll}
    \toprule
    $\ycut$ & $0.01$ & $0.05$ \\
    \midrule
    $\overline{\Gamma}^{\bb{}}(\mathrm{2jet})~[\mev]$ & $1.60395(4)$ & $2.13710(2)$ \\
    $\overline{\Gamma}^{\bb{}}(\mathrm{3jet})~[\mev]$ & $0.75917(4)$ & $0.28118(2)$ \\
    $\overline{\Gamma}^{\bb{}}(\mathrm{4jet})~[\mev]$ & $0.05747$    & $0.0023$     \\
    \bottomrule
  \end{tabular}
  \caption{The total jet rates at NNLO for $\muR=\mh$ using the Durham
    clustering algorithm with $\ycut = 0.01$ and $\ycut = 0.05$,
    including the $\yb\yt$ contribution.}
  \label{tab:jet-rates-msbar-total}
\end{table}
Finally, in \cref{tab:jet-rates-msbar-total} we present the total jet
rates for the central choice of the renormalisation scale,
$\muR=\mh$. We see a clear hierarchy of the jet rates, with the
two-jet rate being the largest and the four-jet rate the smallest for
both clustering parameters. This is more pronounced for the larger jet
cut, $\ycut=0.05$, as expected.

To conclude, we note that the main objective of this section was to
put our NNLO calculation under a thorough examination.
We performed a series of comparisons of our calculation against
various results available in the literature. The positive outcome of
these checks assures us of the validity of our approach.

% -------------------------
\section{Conclusions}
\label{sec:summary}

In this paper, we presented an independent calculation of the NNLO QCD
corrections to the Higgs boson decay into massive $b$-quarks.
We worked in the framework of the nested soft-collinear subtraction
scheme introduced in
Refs.~\cite{Caola:2017dug,Caola:2017xuq,Caola:2019nzf,Caola:2019pfz}.

A complete discussion of all necessary NNLO contributions was
presented in \cref{sec:nnlo}.
In particular, we demonstrated cancellation of all $\ep^{-1}$ poles,
related to soft and collinear singularities of QCD amplitudes.
The cancellation was obtained pointwise in phase space, without
referring to a specific form of the matrix elements.
Furthermore, a full treatment of the $b$-quark mass allowed for an
inclusion of the additional contribution which originates from a
direct interaction of top quarks with the Higgs boson, at a
differential level.

%%%
Our fully differential calculation was implemented in a computer
program.
We carefully tested it by performing a number of cross-checks with
results available in the literature both for the total decay
width~\cite{Harlander:1997xa,Bernreuther:2018ynm} and jet
rates~\cite{Bernreuther:2018ynm}.

%%%
We note that the calculation presented in this paper is an important
step towards a broader phenomenological goal, namely, combining a
description of Higgs boson production with its decay into massive
$b$-quarks.
In the context of associated Higgs production, where large radiative
corrections for important observables have been
reported~\cite{Ferrera:2017zex,Caola:2017xuq,Gauld:2019yng}, a
thorough study of $b$-quark mass effects is an interesting topic for
future research.

% -------------------------
\section*{Acknowledgements}
We wish to thank Kirill Melnikov for stimulating discussions on the
topics treated here and very valuable comments on the manuscript.
We would like to thank Peter Marquard and Matthias Steinhauser for
useful conversations about renormalisation and decoupling.
We are also grateful to the authors of Ref.~\cite{Bernreuther:2018ynm}
for useful correspondence regarding the numerical cross-checks.
This research was supported by the Deutsche Forschungsgemeinschaft
(DFG, German Research Foundation) under grant  396021762 - TRR 257.
The work of A.B. was in part supported by BMBF grant 05H18VKCC1.

% -------------------------
\appendix
\section{Renormalisation}
\label{sec:renorm}
The subtraction scheme that we apply in this calculation is formulated
in terms of UV-renormalised amplitudes. Here, we describe the
renormalisation prescriptions used in our calculation.

We employ a hybrid scheme in which we renormalise the quark and gluon
fields, the quark masses and the Yukawa coupling in an on-shell
scheme, whereas we use the $\MSbar{}$ scheme with five active flavours
for the strong coupling constant. We write
\begin{align}
  \asb        &= (\muR^2)^\ep S_\ep Z_{\as} \as \,, &
  \psi_b^0    &= \sqrt{Z_\psi} \psi_b \,, &
  G_\mu^{0,a} &= \sqrt{Z_A} G_\mu^a \,, &
  m_b^0       &= Z_m m_b \,, &
  y_b^0       &= Z_m y_b \,,
\end{align}
where
$S_\ep = (4 \pi)^{-\ep} \exp(\ep \gamma_\mathrm{E})$.\footnote{This
  agrees with the renormalisation scheme in \cite{Ablinger:2017hst},
  but differs from \cite{Bernreuther:2005gw} where a different choice
  for the renormalisation of the strong coupling is used. Both schemes
  yield the same result for physical observables, but differ by terms
  proportional to $\zeta_2$ in IR-divergent intermediate steps; this
  difference has already been discussed in
  Ref.~\cite{Bernreuther:2018ynm}.}
Due to the choice to enforce the relation
$\yb = \mb (2\sqrt{2} G_F)^{1/2}$, through the renormalisation
condition of the Yukawa coupling, the mass and Yukawa coupling share
the same renormalisation constant. The remaining renormalisation
constants (for the massless quarks, ghosts and the gauge parameter) do
not explicitly appear in our calculation.  The renormalisation of the
strong coupling constant requires \cite{Gross:1973id,Politzer:1973fx}
\begin{align}
  Z_{\as}
    &= 1 + \frac{\as}{4 \pi} Z_{\as}^{(1)} + \order{\as^2}
     = 1 - \frac{\as}{4 \pi} \frac{\beta_0(\nl+1)}{\ep} + \order{\as^2}
  \,,
\end{align}
with $\beta_0(\nl+1) = \frac{11}{3} \CA - \frac{4}{3} \TF (\nl+1)$. The mass and
field renormalisation constants for the massive quark, $Z_\psi$ and $Z_m$, are
given by \cite{Melnikov:2000zc}
%%%texparser:start:ZmZpsi%%%
\begin{align}
%%%texparser:LHS:Zm%%%
  Z_m ={}&
  %%%texparser:stop%%%
      1 +\frac{\asb}{4\pi} Z_m^{(1)}
        +\left(\frac{\asb}{4 \pi}\right)^2 Z_m^{(2)}
        +\order{\asb^3}
  \nonumber \\
    ={}&
  %%%texparser:start:ZmZpsi%%%
      1 +\frac{\asb}{4\pi} (m^2)^{-\ep} S_\ep^{-1}
         \CF \left(- \frac{3}{\ep} - 4 - \ep \left(8 + \frac{3}{2} \zeta_2\right)\right)
  \notag \\ &
        +\left(\frac{\asb}{4\pi}\right)^2 (m^2)^{-2\ep} S_\ep^{-2} \left[
          \CF^2 \left(\frac{9}{2 \ep^2} + \frac{45}{4 \ep} + \frac{199}{8} + \left(-\frac{51}{2} + 48 \ln(2)\right) \zeta_2 - 12 \zeta_3\right)
  \right. \notag \\ &
          +\CA \CF \left(- \frac{11}{2 \ep^2} - \frac{91}{4 \ep} -\frac{605}{8} + \left(\frac{5}{2} - 24 \ln(2)\right) \zeta_2 + 6 \zeta_3\right)
  \notag \\ & \left.
          +\CF \TF \left(\frac{2}{\ep^2} + \frac{7}{\ep} + \frac{69}{2} - 14 \zeta_2\right)
          +\CF \TF \nl \left(\frac{2}{\ep^2} + \frac{7}{\ep} + \frac{45}{2} + 10 \zeta_2\right)
        \right]
        +\order{\asb^3}
  \,, \\
%%%texparser:LHS:Zpsi%%%
  Z_\psi ={}&
  %%%texparser:stop%%%
      1 +\frac{\asb}{4\pi} Z_\psi^{(1)}
        +\left(\frac{\asb}{4 \pi}\right)^2 Z_\psi^{(2)}
        +\order{\asb^3}
  \nonumber \\
    ={}&
  %%%texparser:start:ZmZpsi%%%
      1
      +\frac{\asb}{4\pi} (m^2)^{-\ep} S_\ep^{-1}
        \CF \left(-\frac{3}{\ep} - 4 - \ep \left(8 + \frac{3}{2} \zeta_2\right)\right)
  \notag \\ &
      +\left(\frac{\asb}{4\pi}\right)^2 (m^2)^{-2\ep} S_\ep^{-2} \left[
        \CF^2 \left(\frac{9}{2 \ep^2} + \frac{51}{4 \ep} + \frac{433}{8} + \left(-\frac{147}{2} + 96 \ln(2)\right) \zeta_2 - 24 \zeta_3\right)
  \right. \notag \\ &
        +\CA \CF \left(-\frac{11}{2 \ep^2} - \frac{101}{4 \ep} - \frac{803}{8} + \left(\frac{49}{2} - 48 \ln(2)\right) \zeta_2 + 12 \zeta_3\right)
  \notag \\ & \left.
        +\CF \TF \left(\frac{4}{\ep^2} + \frac{19}{3 \ep} + \frac{1139}{18} - 28 \zeta_2\right)
        +\CF \TF \nl \left(\frac{2}{\ep^2} + \frac{9}{\ep} + \frac{59}{2} + 10 \zeta_2\right)
      \right]
      +\order{\asb^3}
  \,.
\end{align}
%%%texparser:stop%%%
The gluon field renormalisation constant is non-vanishing in the
on-shell scheme due to heavy-quark loops and
reads~\cite{Beenakker:2002nc,Czakon:2007wk}
\begin{align}
  Z_A &= 1 + \frac{\asb}{4 \pi} Z_A^{(1)} + \order{\asb^2}
       = 1 + \frac{\asb}{4 \pi} \frac{ \beta_{0,Q} }{ \ep }
         \left(\frac{\muR^2}{m_b^2}\right)^\ep \Gamma(1+\ep) e^{\ep\EulerGamma}
         +\order{\asb^2}
  \,,
\end{align}
with $\beta_{0,Q} = -\frac{4}{3} \TF$.

For single-virtual and double-virtual amplitudes of the \hbb{} process
we start with the unrenormalised amplitude calculated from all
one-particle-irreducible Feynman diagrams. Expanded in terms of the
bare strong coupling, we have
\begin{align}
  |\KetAmplBare{}{\bb{}}
    &= |\KetAmplBare{0}{\bb{}}
       +\frac{\asb}{4 \pi} |\KetAmplBare{1}{\bb{}}
       +\left(\frac{\asb}{4 \pi}\right)^2 |\KetAmplBare{2}{\bb{}}
       +\order{\asb^3}
\end{align}
and together with the LSZ factors we obtain the renormalised amplitude as
\begin{align}
  |\KetAmpl{}{\bb{}}
    ={}& Z_m Z_\psi |\KetAmplBare{}{\bb{}}
         +\left(\frac{\asb}{4 \pi}\right)^2 Z_m^{(1)} |\KetCT{1}{m,\bb{}}
         +\order{\asb^3}
         \nonumber
  \label{eq:renoHbb}
  \\
    ={}& |\KetAmplBare{0}{\bb{}} 
         +\frac{\as}{4 \pi} (\muR^2)^\ep S_\ep \left[
            |\KetAmplBare{1}{\bb{}}
           +|\KetAmplBare{0}{\bb{}} (Z_m^{(1)} + Z_{\psi}^{(1)})
         \right]
  \notag \\ &
         +\left(\frac{\as}{4 \pi}\right)^2 \left[
           (\muR^2)^\ep S_\ep Z_{\as}^{(1)} \left(
              |\KetAmplBare{1}{\bb{}}
             +|\KetAmplBare{0}{\bb{}} (Z_m^{(1)} + Z_{\psi}^{(1)})
           \right)
  \right. \notag \\ &
           +(\muR^2)^{2\ep} S_\ep^2 \left(
             |\KetAmplBare{2}{\bb{}}
             +(Z_m^{(1)} + Z_{\psi}^{(1)}) |\KetAmplBare{1}{\bb{}}
  \right. \notag \\ & \left.\left.
             +(Z_m^{(2)} + Z_m^{(1)} Z_{\psi}^{(1)} + Z_{\psi}^{(2)})
                |\KetAmplBare{0}{\bb{}}
             +Z_m^{(1)} |\KetCT{1}{m,\bb{}}
           \right)
           \right]
           +\order{\as^3}
           \nonumber
  \\
    ={}& |\KetAmpl{0}{\bb{}} 
         +\frac{\as}{4 \pi} |\KetAmpl{1}{\bb{}}
         +\left(\frac{\as}{4 \pi}\right)^2 |\KetAmpl{2}{\bb{}}
         +\order{\as^3}
  \,.
\end{align}
Here, $|\KetCT{1}{m,\bb{}}$ denotes the amplitude of the mass
counterterm diagrams and $|\KetAmpl{k}{\bb{}}$ denotes the renormalised
amplitude of order $\as^k$.

In contrast to the Born process, the real-emission amplitude for
$\hbb{g}$ starts at $\order{\as}$, which means that coupling
renormalisation already affects the one-loop correction.
In accordance with \cref{eq:Z_op_expansion} we leave powers of the
strong coupling constant related to real emissions implicit in the
amplitude and write the perturbative expansion of the bare $\hbb{g}$
amplitude as
\begin{align}
  |\KetAmplBare{}{\bb{g}}
    &= |\KetAmplBare{0}{\bb{g}}
       +\frac{\asb}{4 \pi} |\KetAmplBare{1}{\bb{g}}
       +\order{\asb^3}
  \,,
\end{align}
In analogy to \cref{eq:renoHbb}, the $\hbb{g}$ process is
renormalised via
\begin{align}
  |\KetAmpl{}{\bb{g}}
    ={}& Z_m Z_\psi \sqrt{Z_A} |\KetAmplBare{}{\bb{g}}
         +\frac{\asb}{4 \pi} Z_m^{(1)} |\KetCT{1}{m,\bb{g}}
         +\order{\asb^3}
         \nonumber
  \label{eq:renoHbbg}
  \\
    ={}& (\muR^2)^\ep S_\ep |\KetAmplBare{0}{\bb{g}}
         +\frac{\as}{4 \pi} \left[
            (\muR^2)^\ep S_\ep Z_{\as}^{(1)} |\KetAmplBare{0}{\bb{g}}
  \right. \notag \\ & \left.
           +(\muR^2)^{2\ep} S_\ep^2 \left(
            |\KetAmplBare{1}{\bb{g}}
            +\left(\frac{Z_{A}^{(1)}}{2} + Z_m^{(1)} + Z_{\psi}^{(1)}\right) |\KetAmplBare{0}{\bb{g}}
            +Z_m^{(1)} |\KetCT{1}{m,\bb{g}}
           \right)
           \right]
           +\order{\as^3}
           \nonumber
  \\
    ={}& (\muR^2)^\ep S_\ep \left(
           |\KetAmpl{0}{\bb{g}} 
           +\frac{\as}{4 \pi} |\KetAmpl{1}{\bb{g}}
         \right)
         +\order{\as^3}
  \,.
\end{align}
Note that for the expansion of the renormalised amplitude, we pull out
a global factor of $(\muR^2)^\ep S_\ep$ and we move it to the
normalisation of the phase-space measure, see \cref{eq:LISP_parton}.
Analogously, the renormalised double-real amplitudes are written as
\begin{align}
  |\KetAmpl{}{\bb{gg}}
    &= (\muR^2)^{2\ep} S_\ep^2
       |\KetAmpl{0}{\bb{gg}}
       +\order{\as^3}
  \,, \\
  |\KetAmpl{}{\bb{\qq}}
    &= (\muR^2)^{2\ep} S_\ep^2
       |\KetAmpl{0}{\bb{\qq}}
       +\order{\as^3}
  \,.
\end{align}

Both for the calculation of the amplitudes and the application of the
subtraction scheme, we choose to work in a theory with $\nf = \nl + 1$
active flavours. This amounts to renormalising the strong coupling
with $\beta_0(\nl + 1)$ and the gluon field with a non-trivial
renormalisation constant ($\sqrt{Z_A} \neq 1$). This procedure takes
care of insertions of heavy-quark loops on gluon into propagators and
external gluon legs.

A possible alternative~\cite{Czakon:2014oma} is to formulate the
subtraction scheme in a theory where the strong coupling evolves with
$\nl$ active flavours. If we calculate the amplitudes in the
renormalisation scheme with $\nl+1$ active flavours described above,
we can use the decoupling relation
\cite{Weinberg:1980wa,Ovrut:1980dg,Wetzel:1981qg,%
  Bernreuther:1981sg,Bernreuther:1983zp,Bernreuther:1983ds,Chetyrkin:1997un}
\begin{align}
  \as^{(\nl)} &= \zeta_{\as} \as^{(\nl+1)}
  \label{eq:as-decoupling}
\end{align}
to absorb the effects of the heavy-quark loops into the running of $\as$.
We stress that in this case, the higher-order terms in $\ep$ of the
decoupling constant $\zeta_{\as}$ need to be taken into account. To
$\order{\as}$ we need \cite{Gerlach:2019kfo}
\begin{align}
  \zeta_{\as}
    &= 1 + \frac{\as^{(\nl)}}{4 \pi} \frac{\beta_{0,Q}}{\ep} \left[
         \left(\frac{\muR^2}{\mb^2}\right)^{\ep} \Gamma(1+\ep)
         e^{\ep \EulerGamma} - 1\right]
         +\order{(\as^{(\nl)})^2}
  \,.
\end{align}
The subtraction scheme then operates on amplitudes in a theory with
$\nl$ active flavours. Finally, after all IR poles cancel we can use
\cref{eq:as-decoupling} in the opposite direction to go back to the
theory with $\nl+1$ active flavours. The results of this procedure are
identical to the results obtained when working with $\nl+1$ active
flavours throughout the whole calculation.

\section{Useful formulae}
For the convenience of the reader, in this appendix, we collect useful
formulae that are available in the literature and are used in our
calculation.

\subsection{Anomalous dimensions for IR factorisation}
\label{sec:Zcoeff}

Below, we give explicit expressions for the anomalous dimensions that
appear in the $\Zop{}{}$ operator. All of these coefficients were
taken from Ref.~\cite{Czakon:2014oma}; they can further be traced to
Refs.~\cite{Becher:2009kw,Becher:2009cu}.

%%%
We write an expansion of the anomalous dimensions in terms of the
strong coupling constant,
\begin{align}
  \gamma_{i}(\as)
  ={}&
       \sum_{n=0}^{\infty} \gamma_{i}^{(n)} \asOnFourPi{}^{n+1}
       \,,
\end{align}
where $i$ stands for the type of the anomalous dimension. We report
formulae for the massless and massive cusp anomalous dimensions,
heavy-quark and gluon anomalous dimensions.

%%%
The expansion coefficients of the massless cusp anomalous dimension are given by
%%%texparser:start:Zop%%%
\begin{align}
  %%%texparser:LHS:gcusp0%%%
  \gamcusp{0} &= 4
  \,, \texparserEOL &
  %%%texparser:LHS:gcusp1%%%
  \gamcusp{1} &= \left(\frac{268}{9} - \frac{4}{3} \pi^2\right) \CA - \frac{80}{9} \TF \nl
  \,,
\end{align}
%%%texparser:stop%%%
with $\nl$ being a number of massless quark flavours.
For the cusp anomalous dimension of massive emitters the expansion
coefficients depend on
\begin{align}
  \label{eq:vij-def}
  v_{ij} &= \sqrt{1-\frac{m_i^2 m_j^2}{\scprod{q_i}{q_j}^2}}
\end{align}
and read
%%%texparser:start:Zop%%%
\begin{align}
  %%%texparser:LHS:gcuspQ0%%%
  \gamcuspQ{0}{v} ={}& \gamcusp{0} \frac{1}{v} \left[\frac{1}{2} \ln\left(\frac{1+v}{1-v}\right) - i\pi\right]
  \,, \\
  %%%texparser:LHS:gcuspQ1%%%
  \gamcuspQ{1}{v} ={}& \gamcusp{1} \frac{1}{v} \left[\frac{1}{2} \ln\left(\frac{1+v}{1-v}\right) - i\pi\right]
    +8 \CA \Bigg\{\zeta_3 - \frac{5}{6} \pi^2 + \frac{1}{4}\ln^2\left(\frac{1+v}{1-v}\right)
  \notag \\ &
    +\frac{1}{v^2}\left[
      \frac{1}{24} \ln^3\left(\frac{1+v}{1-v}\right)
      +\ln\left(\frac{1+v}{1-v}\right) \left(
        \frac{1}{2}\Li_2\left(\frac{1-v}{1+v}\right)
        -\frac{5\pi^2}{12}
      \right)
  \right. \notag \\ & \left.
      +\Li_3\left(\frac{1-v}{1+v}\right)
      -\zeta_3
    \right]
    +\frac{1}{v} \left[
       \frac{5}{6} \pi^2
       +\frac{5}{12} \pi^2 \ln\left(\frac{1+v}{1-v}\right)
       -\frac{1}{4} \ln^2\left(\frac{1+v}{1-v}\right)
  \right. \notag \\ &  \left.
       -\ln\left(\frac{2v}{1+v}\right) \ln\left(\frac{1+v}{1-v}\right)
       -\frac{1}{24} \ln^3\left(\frac{1+v}{1-v}\right)
       +\Li_2\left(\frac{1-v}{1+v}\right)
    \right]
  \notag \\ &
    +i\pi \left\{
      \frac{1}{v^2} \left[
        \frac{\pi^2}{6}
        -\frac{1}{4} \ln^2\left(\frac{1+v}{1-v}\right)
        -\Li_2\left(\frac{1-v}{1+v}\right)
      \right]
  \right. \notag \\ & \left. 
      +\frac{1}{v} \left[
        -\frac{\pi^2}{6}
        +2 \ln\left(\frac{2v}{1+v}\right)
        +\ln\left(\frac{1+v}{1-v}\right)
        +\frac{1}{4} \ln^2\left(\frac{1+v}{1-v}\right)
      \right]
  \right. \notag \\ & \left.
      -\ln\left(\frac{1+v}{1-v}\right)
    \right\}
  \Bigg\}
\end{align}
Furthermore, the heavy-quark anomalous dimensions are
\begin{align}
  %%%texparser:LHS:gQ0%%%
  \gamQ{0} &= -2 \CF
  \,, \texparserEOL &
  %%%texparser:LHS:gQ1%%%
  \gamQ{1} &= \CF \CA \left(\frac{2}{3} \pi^2 - \frac{98}{9} - 4\zeta_3\right)
              +\frac{40}{9} \CF \TF \nl
  \,,
\end{align}
while for gluons we have
\begin{align}
  %%%texparser:LHS:gg0%%%
  \gamg{0} &=
  %%%texparser:stop%%%
              -\beta_0(\nl)
            =
  %%%texparser:start:Zop%%%
              -\frac{11}{3} \CA + \frac{4}{3} \TF \nl
  \,.
\end{align}
%%%texparser:stop%%%

\subsection{Coefficients for on-shell to \texorpdfstring{\MSbar{}}{MSbar}-scheme conversion}
\label{sec:pole2msbar-coeff}

The coefficients for the conversion relation between the on-shell and
\MSbar{} Yukawa coupling, defined in \cref{eq:yukawa-conversion}, are
given by~\cite{Melnikov:2000qh,Bernreuther:2018ynm}
%%%texparser:start:pole2msbar%%%
\begin{align}
  %%%texparser:LHS:pole2msbarR1%%%
  r_1 ={}& -2d_1\,, & \texparserEOL{}
  %%%texparser:LHS:pole2msbarR2%%%
  r_2 ={}& 3d_1^{\,2} - 2d_2\,,
\end{align}
and
\begin{align}
  %%%texparser:LHS:pole2msbarD1%%%
  d_1(\mb,\mu)
  ={}&
       - \CF \lp{1 + \frac{3}{4}L}\rp
       \,,
  \\
  %%%texparser:LHS:pole2msbarD2%%%
  d_2(\mb,\mu)
  ={}&
       \CF^2 \lb{
       \frac{7}{128}
       - \frac{3}{4}{\zeta_3}
       + 3\ln(2){\zeta_2}
       - \frac{15}{8}{\zeta_2}
       + \frac{21}{32}L + \frac{9}{32}L^2
       }\rb
       \nonumber\\
     &
       +\CA \CF\lb{
       -\frac{1111}{384}
       + \frac{3}{8}{\zeta_3}
       + \frac{1}{2}{\zeta_2}
       - \frac{3}{2}\ln(2){\zeta_2}
       - \frac{185}{96}L - \frac{11}{32}L^2
       }\rb
       \nonumber\\
     &
       +\CF \TF \nl\lb{
       \frac{71}{96}
       + \frac{1}{2}{\zeta_2}
       + \frac{13}{24}L + \frac{1}{8}L^2
       }\rb
       %%%
       +\CF \TF\lb{
       \frac{143}{96}
       - {\zeta_2}
       + \frac{13}{24}L + \frac{1}{8}L^2
       }\rb
       \,,
\end{align}
%%%texparser:stop%%%
with the abbreviation $L = \ln\lp{ \mu^2/\mb^2 }\rp$.

\section{Factorisation formulae}
\label{sec:fact-form}
Here, we collect the factorisation formulae which are necessary to
evaluate the singular limits of the squared matrix elements. We
specialise to the $\hbb{g}$, $\hbb{gg}$ and $\hbb{\qq}$ processes.
For a useful collection of factorisation formulae for general NNLO QCD
processes, we refer the reader to Ref.~\cite{Czakon:2014oma}.

\subsection{Single-collinear factorisation}
\label{sec:fact-single-coll}

In order to discuss the single-collinear limit of the \hbb{gg} and
\hbb{\qq} matrix element, we have to define a perpendicular direction
$\ktmu$ which determines how the collinear limit is approached. We
choose
\begin{align}
  \ktmu
  ={}&
            \lim_{\eta\to0}
            \frac{ \hat{q}_5 - \hat{q}_4 }{ \| \hat{q}_5 - \hat{q}_4  \| }
  \,,
\end{align}
where $\eta$ refers to the phase-space parametrisation in
\cref{eq:phsp_hbbgg}.
Then the factorisation formula in the single-collinear limit
reads
\begin{align}
  \label{eq:coll-fact}
  \coll{45} \MEsq{0}{\bb{ij}}
  ={}&
       \frac{ \gs^2 }{ \scprod{q_4}{q_5} }
       \BraAmpl{0}{\bb{g},\mu}|
       \Pij{0}{ij}{\mu\nu}(z,k_\perp;\ep)
       |\KetAmpl{0}{\bb{g},\nu}\,,
\end{align}
where $ij$ stands for either the $gg$ or $\qq$ channel and
$|\KetAmpl{0}{\bb{g},\nu}$ is the spin-correlated $\hbb{g}$ amplitude
with the gluon polarisation vector removed, i.e.
\begin{align}
  |\KetAmpl{0}{\bb{g}}
  ={}&
       (\varepsilon^{\mu}(q_4,\lambda))^{\ast}
       |\KetAmpl{0}{\bb{g},\mu}
       \,.
\end{align}
The splitting functions $\Pij{0}{ij}{\mu\nu}$ carry Lorentz indices
which are contracted with the corresponding index of the gluon. For the
$gg$ and $\qq{}$ channels they read
\begin{align}
  \label{eq:split-Pgg}
  \Pij{0}{gg}{\mu\nu}(z,k_\perp;\ep)
  ={}&
       2\CA
       \lb{
       -\gmn \lp{ \frac{z}{1-z} + \frac{1-z}{z}}\rp
       -2(1-\ep)z(1-z)\frac{ \ktmu\ktnu }{ \ktsq }
       }\rb\,,
  \\
  \label{eq:split-Pqq}
  \Pij{0}{\qq}{\mu\nu}(z,k_\perp;\ep)
  ={}&
       \TF
       \lb{
       -\gmn
       +4z(1-z)\frac{ \ktmu\ktnu }{ \ktsq }
       }\rb\,,
\end{align}
with
\begin{align}
  \label{eq:z-coll}
  z ={}& \frac{E_4}{E_4 + E_5} = 1 - \frac{\xi_2}{2} \,.
\end{align}

As the factorisation formula above indicates, we have to evaluate
spin-correlated matrix elements. In the term where they are contracted
with the metric tensor $\gmn$, the result corresponds to an
uncorrelated matrix element via the polarisation sum. This leads to
\begin{align}
  \label{eq:polarisation-sum-me}
  \BraAmpl{0}{\bb{g},\mu}|
  (-\gmn)
  |\KetAmpl{0}{\bb{g},\nu}
  ={}&
       \MEsq{0}{\bb{g}}
       \,.
\end{align}

For four-dimensional matrix elements, the product
$\ktmu |\KetAmpl{0}{\bb{g},\nu}$ can be linked to helicity amplitudes
using the fact that the perpendicular vector $\ktmu$ may be decomposed
as
\begin{align}
  \ktmu
  ={}&
       - \sum_{\lambda}
       \scprod{\varepsilon(q_4,\lambda)}{k_\perp}
       (\varepsilon^{\mu}(q_4,\lambda))^{\ast}
       \,.
\end{align}
This is true since, by construction, $\ktmu$ lies in the plane
perpendicular to the momentum of a gluon, $q_4$, which is spanned by
the polarisation vectors $\varepsilon^{\mu}(q_4,\lambda)$.

\subsection{Single-soft factorisation (tree-level)}
\label{sec:fact-single-soft}

In the single-soft limits, the factorisation formulae for the $\hbb{g}$ and
$\hbb{gg}$ squared matrix elements are given by
\begin{align}
  \label{eq:soft-fact}
  \soft{4} \MEsq{0}{\bb{g}}
  ={}&
     -\gs^{\,2} \,
     \CF\lp{ \Sij{0}{22}{4} - 2\Sij{0}{23}{4} + \Sij{0}{33}{4} }\rp
     \MEsq{0}{\bb{}}
  \,,
  \\
  \label{eq:single-soft-hbbgg}
  \soft{5} \MEsq{0}{\bb{gg}}
  ={}&
       -\gs^{\,2} \,
       \lb{
        \CF\lp{ \Sij{0}{22}{5} - 2\Sij{0}{23}{5} + \Sij{0}{33}{5} }\rp
       -\CA\lp{ \Sij{0}{24}{5} + \Sij{0}{34}{5} - \Sij{0}{23}{5} }\rp
       }\rb
       \MEsq{0}{\bb{g}}
  \,,
\end{align}
where $\Sij{0}{ij}{k}$ is the usual tree-level eikonal factor
\begin{align}
  \label{eq:single-eikonal}
  \Sij{0}{ij}{k}
  ={}&
       \frac{ \scprod{q_i}{q_j} }{ \scprod{q_i}{q_k}\scprod{q_j}{q_k} }\,.
\end{align}

\subsection{Single-soft factorisation (one-loop)}
\label{sec:fact-single-soft-1l}

The soft limit of the one-loop amplitudes with massive quarks has been
studied in Refs.~\cite{Mitov:2006xs,Bierenbaum:2011gg}. In our case,
it can be written as
\begin{align}
  \label{eq:soft-fact-1loop}
  \soft{4} \, 2\Re\BraAmpl{0}{\bb{g}} | \KetAmpl{1}{\bb{g}}
  ={}&
       -\gs^{\,2}
       \CF
       \lp{ \Sij{0}{22}{4} - 2\Sij{0}{23}{4} + \Sij{0}{33}{4} }\rp
  \nonumber \\ &
       \times \lb{
       2 \Re \BraAmpl{0}{\bb{}} | \KetAmpl{1}{\bb{}}
       +\lp{\Rij{1}{23}{4} + Z_{\as}^{(1)} + Z_A^{(1)} }\rp
       \MEsq{0}{\bb{}}
       }\rb
  \,,
\end{align}
where in the second line we single out the contribution that comes
from the renormalisation procedure, i.e. it involves terms resulting
from the strong-coupling renormalisation, $Z_{\as}^{(1)}$, as well as a
term from the gluon wave-function renormalisation, $Z_A^{(1)}$, see
\cref{sec:renorm}.
Furthermore, the functions $\Rij{1}{ij}{4}$ denote the one-loop eikonal
factor that can be expanded in an $\ep$ series as
\begin{align}
  \label{eq:Rij-def}
  \Rij{1}{ij}{4}
  ={}&
       4\CA
       \lp{ \tfrac{1}{2}\muR^2 \Sij{0}{ij}{4} }\rp^{\ep}
       \sum_{k=-2}^{1}
       \ep^k R_{k}(q_i,q_j;q_4)\,,
\end{align}
where functions $R_{k}(q_i,q_j;q_4)$ have been calculated in
Ref.~\cite{Bierenbaum:2011gg}, and further simplified in
Ref.~\cite{Czakon:2018iev}. In our case we use the formulae from
Eq.~(4) of Ref.~\cite{Czakon:2018iev}; we emphasise that the
expressions therein correspond to the unrenormalised one-loop
soft-gluon current. This is particularly convenient, since we perform
the renormalisation in a hybrid scheme, as outlined in
\cref{sec:renorm}.

Finally, we note that, for our calculation, we split the one-loop matrix
elements in the factorisation formula, \cref{eq:soft-fact-1loop}, into
finite terms and terms containing poles in $\ep^{-1}$ using the
$\Zop{}{}$ operator. This is reflected in \cref{eq:RVF,eq:RVSU}.

\subsection{Double-soft factorisation}
\label{sec:fact-double-soft}

The relevant factorisation formulae for amplitudes that involve massive
particles can be obtained from Ref.~\cite{Catani:1999ss,Czakon:2011ve},
see also Ref.~\cite{Czakon:2014oma}.
In general, the double-soft limit
requires single- and double-eikonal factors and colour-correlated matrix
elements.
For the $\hbb{gg}$ and \hbb{\qq} matrix elements, the colour-correlated
matrix elements can be expressed explicitly through $\mathrm{SU}(3)$
colour factors so that we arrive at
\begin{align}
  \label{eq:dble_soft_ij}
  \dsoft{45} \MEsq{0}{\bb{ij}}
  ={}&
       \gs^{\,4} \, \DSoft{0}{ij}(q_2,q_3;q_4,q_5) \MEsq{0}{\bb{}}
  \,,
\end{align}
where $\DSoft{0}{ij}(q_2,q_3;q_4,q_5)$ denotes the double-soft functions
for the partons $ij \in \{gg,\qq{}\}$. They are given by
\begin{align}
  \label{eq:dble_soft_gg_def}
  \DSoft{0}{gg}(q_2,q_3;q_4,q_5)
  ={}&
       \CF^2
       \lp{ \Sij{0}{22}{4} - 2\Sij{0}{23}{4} + \Sij{0}{33}{4} }\rp
       \lp{ \Sij{0}{22}{5} - 2\Sij{0}{23}{5} + \Sij{0}{33}{5} }\rp
       \nonumber\\
  &
       -\CA\CF\
       \lp{ \SSij{gg}{22}{45} - \SSij{gg}{23}{45} - \SSij{gg}{32}{45} + \SSij{gg}{33}{45} }\rp
  \,,
  \\
  \label{eq:dble_soft_qq_def}
  \DSoft{0}{\qq}(q_2,q_3;q_4,q_5)
  ={}&
       \CF\TF
       \lp{
       \SSij{\qq{}}{22}{45} - \SSij{\qq{}}{23}{45}
       - \SSij{\qq{}}{32}{45} + \SSij{\qq{}}{33}{45} }
       \rp\,,
\end{align}
where $\Sij{0}{ij}{k}$ is the usual single-eikonal factor, defined in
\cref{eq:single-eikonal}, and $\SSij{gg}{ij}{45}$ and
$\SSij{\qq}{ij}{45}$ are the double-eikonal factors. For the $gg$
emission case, we have
\begin{align}
  \SSij{gg}{ij}{45}
  ={}&
       \SSij{(m=0)}{ij}{45}
       + m_i^{\,2}\SSij{(m\neq0)}{ij}{45}
       + m_j^{\,2}\SSij{(m\neq0)}{ji}{45}\,,
\end{align}
where~\cite{Catani:1999ss}
%%%texparser:start:dsoft%%%
\begin{align}
  %%%texparser:LHS:SSijCoeffm0%%%
  \label{eq:SSij-coeff-massless}
  \SSij{(m=0)}{ij}{45}
  ={}&
       \frac{ (1-\ep) }{ \scprod{q_4}{q_5}^2 }
       \frac
       { \scprod{q_i}{q_4}\scprod{q_j}{q_5} + \scprod{q_i}{q_5}\scprod{q_j}{q_4} }
       { \scprod{q_i}{q_{45}}\scprod{q_j}{q_{45}} }
       \nonumber\\
  &
       -\frac{1}{2}
       \frac{\scprod{q_i}{q_j}}{\scprod{q_i}{q_4}\scprod{q_j}{q_4}}
       \frac{\scprod{q_i}{q_j}}{\scprod{q_i}{q_5}\scprod{q_j}{q_5}}
       \lb{
       2 - \frac
       { \scprod{q_i}{q_4}\scprod{q_j}{q_5} + \scprod{q_i}{q_5}\scprod{q_j}{q_4} }
       { \scprod{q_i}{q_{45}}\scprod{q_j}{q_{45}} }
       }\rb
       \nonumber\\
  &
       +\frac{ 1 }{ \scprod{q_4}{q_5} }
       \lb{
        \frac{ \scprod{q_i}{q_j} }{ \scprod{q_i}{q_4}\scprod{q_j}{q_5} }
       +\frac{ \scprod{q_i}{q_j} }{ \scprod{q_i}{q_5}\scprod{q_j}{q_4} }
       }\right.
       \nonumber\\
  &
  \left.{
       -\frac{ \scprod{q_i}{q_j} }{ \scprod{q_i}{q_{45}}\scprod{q_j}{q_{45}} }
       \lp{
       2 +
       \frac
       { \lb{ \scprod{q_i}{q_4}\scprod{q_j}{q_5}+\scprod{q_i}{q_5}\scprod{q_j}{q_4} }\rb^2 }
       { 2\scprod{q_i}{q_4}\scprod{q_j}{q_5}\scprod{q_i}{q_5}\scprod{q_j}{q_4} }
       }\rp
       }\rb
\end{align}
and~\cite{Czakon:2011ve}
\begin{align}
  %%%texparser:LHS:SSijCoeffm%%%
  \label{eq:SSij-coeff-massive}
  \SSij{(m\neq0)}{ij}{45}
  ={}&
       -\frac{1}{4 \scprod{q_i}{q_4}\scprod{q_i}{q_5}\scprod{q_4}{q_5}}
       +\frac{1}{2}\frac
       { \scprod{q_i}{q_j} }
       { \scprod{q_i}{q_4}\scprod{q_i}{q_5}\scprod{q_j}{q_4}\scprod{q_j}{q_5} }
       \frac{ \scprod{q_j}{q_{45}} }{ \scprod{q_i}{q_{45}} }
       \nonumber\\
  &
       -\frac{1}{2}
       \frac{1}{ \scprod{q_4}{q_5}\scprod{q_i}{q_{45}}\scprod{q_j}{q_{45}} }
       \lp{
       \frac{ \scprod{q_j}{q_4}^2 }{ \scprod{q_i}{q_4}\scprod{q_j}{q_5} }
       +\frac{ \scprod{q_j}{q_5}^2 }{ \scprod{q_i}{q_5}\scprod{q_j}{q_4} }
       }\rp
  \,.
\end{align}
%%%texparser:stop%%%
In \cref{eq:SSij-coeff-massless,eq:SSij-coeff-massive} the shorthand
$q_{45}=(q_4+q_5)$ is used.
%% qq~ soft emissions
For the case of a soft $\qq{}$ pair emission, the double-eikonal factor
is given by \cite{Catani:1999ss}
%%%texparser:start:dsoft%%%
\begin{align}
  %%%texparser:LHS:IIijCoeff%%%
  \SSij{\qq{}}{ij}{45}
  ={}&
       \frac{1}{ \scprod{q_4}{q_5}^2 }
       \frac
       {
       \scprod{q_i}{q_4}\scprod{q_j}{q_5}
       + \scprod{q_i}{q_5}\scprod{q_j}{q_4}
       - \scprod{q_i}{q_j}\scprod{q_4}{q_5}
       }
       { \scprod{q_i}{q_{45}}\scprod{q_j}{q_{45}} }
       \,.
\end{align}
%%%texparser:stop%%%

In the strongly-ordered double-soft limit, $\soft{5}\dsoft{45}$, where
we take both $\xi_1 \to 0$ and $\xi_2 \to 0$, the double-soft function
simplifies to
\begin{align}
  \label{eq:dble-soft-so}
  \soft{5} \DSoft{0}{gg}(q_2,q_3;q_4,q_5)
  ={}&
  \CF^2
  \lp{ \Sij{0}{22}{4} - 2\Sij{0}{23}{4} + \Sij{0}{33}{4} }\rp
  \lp{ \Sij{0}{22}{5} - 2\Sij{0}{23}{5} + \Sij{0}{33}{5} }\rp
  \nonumber\\
  &
  +\CF\CA
  \lp{ \Sij{0}{22}{4} - 2\Sij{0}{23}{4} + \Sij{0}{33}{4} }\rp
  \lp{ \Sij{0}{23}{5} - \Sij{0}{24}{5} - \Sij{0}{34}{5} }\rp
  \,.
\end{align}
The corresponding limit for the case of $\qq$ emission is regular.

Moreover, we need the double-soft single-collinear limit,
$\dsoft{45}\coll{45}$, which can be obtained by taking the collinear
limit of the double-soft function. However, a simpler expression
arises if we use an iterated factorisation formula, taking first the
collinear ($q_4 \parallel q_5$) and then the soft limit of the parent
parton of the splitting ($q_{45}^0 \to 0$). We obtain
\begin{align}
  \dsoft{45}\coll{45} \MEsq{0}{\bb{ij}}
    &= \frac{\gs^4}{\scprod{q_4}{q_5}}
       \Pij{0}{ij}{\mu\nu} \CF \lp{
         \Sij{0}{22}{(45),\mu\nu}
         -2 \Sij{0}{23}{(45),\mu\nu}
         +\Sij{0}{33}{(45),\mu\nu}
       }\rp
       \MEsq{0}{\bb{}}
  \,,
\end{align}
where we use the shorthand notation
\begin{align}
  \Sij{0}{ij}{(45),\mu\nu}
    &= \frac{q_{i,\mu} \, q_{j,\nu}}{\scprod{q_i}{q_{45}} \scprod{q_{45}}{q_j}}
  \,.
\end{align}

\section{Integrated subtraction terms}
\label{sec:subt_int}
In this section we report formulae for integrated subtraction terms
that we use throughout the calculation. For the convenience of
the reader, we also include results available in the literature.

\subsection{Single-collinear subtraction terms}
\label{sec:Pij0int}
The relevant factorisation formula for the single-collinear limit is
given in \cref{eq:coll-fact}. For the integrated subtraction terms, we
integrate the splitting function over the unresolved phase space in $d$
dimensions.

We recall that the splitting functions in
\cref{eq:split-Pgg,eq:split-Pqq} contain a term proportional to
$\ktmu\ktnu/\ktsq$ which is contracted with the spin-correlated
matrix element $\BraAmpl{0}{\bb{g},\mu}|\KetAmpl{0}{\bb{g},\nu}$.
In contrast to the single-collinear subtraction terms, where the spin
correlations are required to make the subtraction local, the integrated
subtraction term can be averaged over the azimuthal directions
of momentum $q_5$. The reduced matrix element depends only on the sum of
momenta, $q_{45} = (q_4 + q_5)$, which is independent of the azimuthal
direction of $q_5$.
Therefore, the integral over $\dOmega{5}{2-2\ep}$ decouples and yields
\begin{align}
  \label{eq:azim-avg}
  \left(\int \dOmega{5}{2-2\ep}\right)^{-1}
  \int \dOmega{5}{2-2\ep}
  \frac{\ktmu\ktnu}{\ktsq}
  ={}&
       \frac{1}{2(1-\ep)}
       \lb{
       \gmn
       -\frac{ n^{\mu}\bar{n}^{\nu} + \bar{n}^{\mu} n^{\nu} }{ \scprod{n}{\bar{n}} }
       }\rb
       \,,
\end{align}
with $\bar{n}^{\mu} = n_{\nu}$ and $n^{\mu} = q_{4}^{\mu} / q_{4}^{0}$
in the collinear limit. To derive this result, we use the fact that
the right-hand side of \cref{eq:azim-avg} is invariant under
rotations in the $(2-2\ep)$-dimensional sphere, that it has to be
orthogonal to $q_4^\mu$ and that by definition the time components
($\mu=0$ and $\nu=0$) need to vanish \cite{Czakon:2014oma}.
Note that, due to the Ward identity $q_4^{\mu} |\KetAmpl{0}{\bb{g},\mu} ={} 0$,
the $n^\mu \bar{n}^\nu$ and $\bar{n}^\mu n^\nu$ terms in \cref{eq:azim-avg}
drop out when contracted with the spin-correlated squared matrix
elements.\footnote{Note that this argument is not necessarily valid if more
  that one gluon is spin-correlated in the reduced amplitude.}

This means that after azimuthal averaging and using
\cref{eq:polarisation-sum-me} we replace
\begin{align}
  \frac{\ktmu\ktnu}{\ktsq}
  \BraAmpl{0}{\bb{g},\mu}|\KetAmpl{0}{\bb{g},\nu}
  \longrightarrow{}
  &
    -\frac{1}{2(1-\ep)}
    \MEsq{0}{\bb{g}}
    \,.
\end{align}
As a result the factorisation formula of \cref{eq:coll-fact} simplifies
to
\begin{align}
  \coll{45} \MEsq{0}{\bb{ij}}
  ={}&
       \frac{ \gs^2 }{ \scprod{q_4}{q_5} }
       \PijAvg{0}{ij}{z;\ep}
       \MEsq{0}{\bb{g}}
       \,,
\end{align}
where $\PijAvg{0}{ij}{z;\ep}$ are the azimuthally averaged splitting
functions and read
\begin{align}
  \PijAvg{0}{gg}{z;\ep}
  ={}&
       2\CA
       \lb{
       \frac{z}{1-z}
       +\frac{1-z}{z}
       +z(1-z)
       }\rb\,,
  \\
  \PijAvg{0}{\qq}{z;\ep}
  ={}&
       \TF
       \lb{
       1
       -\frac{2z(1-z)}{1-\ep}
       }\rb\,.
\end{align}

To obtain the integrated collinear subtraction term, we perform the
integral over the unresolved phase space, $\dq{5}$, using the
parametrisation of \cref{eq:phsp_hbbgg} in the collinear limit.
We arrive at
\begin{align}
  \coll{45} \int\d\Phi_{\bb{gg}}^{E_4 > E_5}(q_1)
  \MEsq{0}{\bb{ij}}
  ={}&
       \gs^2 \PijInt{0}{ij}
       \int\d\Phi_{\bb{g}}(q_1)
       \,
       (\xi_1/2)^{-2\ep}
       \MEsq{0}{\bb{g}}\,,
\end{align}
with the integrated splitting function $\PijInt{0}{ij}$ defined as
\begin{align}
  \label{eq:PijInt-def}
  \PijInt{0}{ij}
  ={}&
       (\muR^2)^{\ep}\Sep
       \Emax^{2-2\ep}
       \lp\frac{\xi_1}{2}\rp^{2}
       \int \frac{\dOmega{5}{2-2\ep}}{2 (2\pi)^{3-2\ep}}
       \int_0^1 \frac{ \d\eta }{ \eta^\ep }
       \int_0^1 \d \xi_2
       (\xi_2(2-\xi_2))^{1-2\ep}
       \frac{ \PijAvg{0}{ij}{1-\tfrac{\xi_2}{2};\ep} }{%
         \scprod{q_4}{q_5} }
  \nonumber \\
  ={}&
       \frac{1}{2}
       (\muR^2)^{\ep}\Sep
       \Emax^{-2\ep}
       \int \frac{\dOmega{5}{2-2\ep}}{2 (2\pi)^{3-2\ep}}
       \int_0^1 \frac{ \d\eta }{ \eta^{1+\ep} }
       \int_0^1 \d \xi_2
       (\xi_2(2-\xi_2))^{-2\ep}
       \PijAvg{0}{ij}{1-\tfrac{\xi_2}{2};\ep}
       \,.
\end{align}

To perform the integral in \cref{eq:PijInt-def} we use the fact
that the integrand is symmetric under
$\xi_2 \leftrightarrow (2-\xi_2)$ exchange and hence we can extend the
integration region to $0< \xi_2 < 2$ at at cost of introducing a
factor of $1/2$. Then we obtain
%%%texparser:start:PijInt%%%
\begin{align}
  \label{eq:PijInt-result-gg}
  %%%texparser:LHS:PijInt0gg%%%
  \PijInt{0}{gg}
  ={}&
       \lp{\frac{\muR^2}{16\Emax^2}}\rp^{\ep}\,
       \frac{ \CA }{ (4\pi)^2 }\,
       \frac{ 6(2-3\ep) }{ \ep^2 }\,
       \frac{ e^{\ep\EulerGamma} \Gamma^{2}(2-2\ep) }{ \Gamma(4-4\ep) \Gamma(1-\ep) }
       \,,
       \\
  \label{eq:PijInt-result-qq}
  %%%texparser:LHS:PijInt0qq%%%
  \PijInt{0}{\qq}
  ={}&
       \lp{\frac{\muR^2}{16\Emax^2}}\rp^{\ep}\,
       \frac{ \TF }{ (4\pi)^2 }\,
       \frac{ \ep(10-8\ep)-4 }{ \ep }\,
       \frac{ e^{\ep\EulerGamma} \Gamma(1-2\ep) \Gamma(2-2\ep) }{ \Gamma(4-4\ep) \Gamma(2-\ep) }
       \,.
\end{align}
In the case of $\PijInt{0}{gg}$, it is necessary to also calculate the
integrated splitting function in the $\xi_2 \to 0$ limit. We find
\begin{align}
  \label{eq:PijInt-soft-result-gg}
  %%%texparser:LHS:PijSoftInt0gg%%%
  \PijSoftInt{0}{gg}
  ={}&
       \lp{\frac{\muR^2}{4\Emax^2}}\rp^{\ep}\,
       \frac{ \CA }{ (4\pi)^2 }\,
       \frac{ 2 }{ \ep^2 }\,
       \frac{ e^{\ep\EulerGamma} }{ \Gamma(1-\ep) }
       \,.
\end{align}
%%%texparser:stop%%%

\subsection{Single-soft subtraction terms (tree-level)}
\label{sec:Sij0int}
When considering a single-soft emission of a gluon with momentum $q_k$
that involves partons with momenta $q_i$ and $q_j$, we encounter the
tree-level eikonal factor $\Sij{0}{ij}{k}$, defined in
\cref{eq:single-eikonal}. In the following we will use the notation
\begin{align}
  \label{eq:qhat-def}
  \hat{q}^\mu &= \frac{q^\mu}{q^0}
\end{align}
to denote a momentum rescaled by its energy component.

The soft function is integrated over the unresolved phase space using
the parametrisation in \cref{eq:phsp_hbbg}, where the Born phase space
decouples in the soft limit.
Thus, the gluon energy is unconstrained unless we insert some bound by
hand. For simplicity, we keep the integration domain of $\xi_1$
unchanged, which corresponds to an upper limit of the energy of $\Emax$.
We arrive at
\begin{align}
  \SijInt{0}{ij}
  ={}&
       \int \dq{4}
       \Sij{0}{ij}{4}
       \nonumber\\
  ={}&
       \lp\frac{\muR^2}{\Emax^2}\rp^\ep \Sep
       \int_0^1 \frac{\d\xi_1}{\xi_1^{1+2\ep}}
       \int\frac{\dOmega{4}{3-2\ep}}{2(2\pi)^{3-2\ep}}
       \frac{\scprod{q_i}{q_j}}{\scprod{q_i}{\hat{q}_4}\scprod{\hat{q}_4}{q_j}}
       \nonumber\\
  ={}&
       \lp\frac{\muR^2}{\Emax^2}\rp^\ep \Sep
       \lb-\frac{1}{2\ep}\rb
       \int\frac{\dOmega{4}{3-2\ep}}{2(2\pi)^{3-2\ep}}
       \frac{\scprod{q_i}{q_j}}{\scprod{q_i}{\hat{q}_4}\scprod{\hat{q}_4}{q_j}}\,,
\end{align}
where the pole in the last line arises from performing the energy
integral. We are left with angular integrals only, for which we write
\begin{align}
  \int\dOmega{4}{3-2\ep}
  \frac{\scprod{q_i}{q_j}}{\scprod{q_i}{\hat{q}_4}\scprod{\hat{q}_4}{q_j}}
  ={}&
       \int\dOmega{4}{1-2\ep}
       \int\d(\cos\theta)
       \d\phi(\sin\theta\sin\phi)^{-2\ep}
       \frac{\scprod{q_i}{q_j}}{\scprod{q_i}{\hat{q}_4}\scprod{\hat{q}_4}{q_j}}
       \nonumber\\
  ={}&
       \frac{\Gamma(1-\ep)}{(4\pi)^{\ep}\Gamma(1-2\ep)} (2\pi) I(q_i,q_j)
  \,.
\end{align}
Here, we introduce the auxiliary function
\begin{align}
  \label{eq:Ipq-def}
  I(q_i,q_j)
  ={}&
       \int\d(\cos\theta)
       \int\frac{\d\phi}{\pi}(\sin\theta\sin\phi)^{-2\ep}
       \frac{\scprod{q_i}{q_j}}{\scprod{q_i}{\hat{q}_4}\scprod{\hat{q}_4}{q_j}}\,.
\end{align}
We write it as a Laurent series in $\ep$, i.e.
\begin{align}
  \label{eq:Ipq-exp}
  I(q_i,q_j)
  ={}&
       \sum_{k=-1}^2 \ep^k I^{(k)}(q_i,q_j)
       + \mathcal{O}(\ep^3)\,,
\end{align}
where the coefficients $I^{(k)}(q_i,q_j)$ depend on the momenta $q_i$
and $q_j$, in particular on whether they are massless or massive. All
necessary coefficients for the single-soft integrated subtraction terms,
except for the $\order{\ep^2}$ terms, have been obtained in
Ref.~\cite{Alioli:2010xd}, see Appendix~A therein. For completeness, we
collect below the formulae for those $I^{(k)}(q_i,q_j)$ which are
relevant to our calculation.

We write the full integrated single-soft eikonal factor as
\begin{align}
  \SijInt{0}{ij}
  ={}&
       - \, \frac{1}{(4\pi)^2} \, \frac{1}{\ep} \, \frac{e^{\ep\EulerGamma} \Gamma(1-\ep)}{\Gamma(1-2\ep)}
       \lp\frac{\muR^2}{4\Emax^2}\rp^\ep
       I(q_i,q_j) \,.
\end{align}
We have to distinguish the case where the emitters are two massive
particles, which we denote by a subscript $\mathrm{MM}$, and the case
where emitter $i$ is massive and emitter $j$ is massless, which we
denote by $\mathrm{M0}$.
For our calculation, we need the coefficients
$I_{\mathrm{MM}}^{(k)}(q_i,q_j)$ and $I_{\mathrm{M0}}^{(k)}(q_i,q_j)$
for $k \in \{-1,0,1\}$ when the directions of the momenta $q_i$ and $q_j$
are arbitrary. Moreover, we need $I_{\mathrm{MM}}^{(k)}(q_i,q_j)$ for
$k \in \{-1,0,1,2\}$ in two special cases: for the case where the two
momenta are equal, $q_i = q_j$ and for the case where $q_i$ and $q_j$
are in a back-to-back configuration.

\paragraph{Two massive emitters}
The reported formulae correspond to the result outlined in
Eqs.~(A.41) to (A.51) of Ref.~\cite{Alioli:2010xd}.
We consider two time-like momenta $q_i^{\,2} = m_i^{\,2}$ and
$q_j^{\,2} = m_j^{\,2}$. With the definition of $v_{ij}$ from
\cref{eq:vij-def} and the notation
\begin{align}
  &\vec{u} ={} \frac{\vec{q}_i}{E_i}\,,&
  &\textrm{ and }&
  &\vec{w} ={} \frac{\vec{q}_j}{E_j}\,,&
\end{align}
we introduce the shorthand notations
\begin{align}
  A^2 ={}&
         \scprod{\vec{u}}{\vec{u}} + \scprod{\vec{w}}{\vec{w}} - 2 \scprod{\vec{u}}{\vec{w}}\,,&
  X_1 ={}&
           \scprod{\vec{u}}{\vec{u}} - \scprod{\vec{u}}{\vec{w}}\,,
         \nonumber\\
  B^2 ={}&
         \scprod{\vec{u}}{\vec{u}}\scprod{\vec{w}}{\vec{w}} -  \scprod{\vec{u}}{\vec{w}}^2\,,&
  X_2 ={}&
           \scprod{\vec{w}}{\vec{w}} - \scprod{\vec{u}}{\vec{w}}\,.
\end{align}
Furthermore, we need the following arguments
\begin{align}
  z_{+} ={}& A + \sqrt{A^2-B^2}\,,&
  z_{1} ={}& \sqrt{ X_1^{\,2} + B^2 } - X_1\,,
           \nonumber\\
  z_{-} ={}& A - \sqrt{A^2-B^2}\,,&
  z_{2} ={}& \sqrt{ X_2^{\,2} + B^2 } + X_2\,,
\end{align}
which will be used in the function
\begin{align}
  K(z)
  ={}&
       - 2\mathrm{Li}_2\lp{\frac{  2z_{-}(z_{+}-z_{}) }{ (z_{+}-z_{-})(z_{-}+z_{}) }}\rp
       - 2\mathrm{Li}_2\lp{\frac{ -2z_{+}(z_{-}+z_{}) }{ (z_{+}-z_{-})(z_{+}-z_{}) }}\rp
       \nonumber\\
     &
       -\frac{1}{2}\log^2\lp{\frac{(z_{}-z_{-})(z_{+}-z_{})}{(z_{}+z_{-})(z_{+}+z_{})}}\rp\,.
\end{align}
With these abbreviations, the coefficients of the massive-massive
angular integral read
\begin{align}
  \label{eq:IMM_general_-1}
  I_{\mathrm{MM}}^{(-1)}(q_i,q_j) ={}& 0\,, \\
  \label{eq:IMM_general_0}
  I_{\mathrm{MM}}^{( 0)}(q_i,q_j)  ={}& \frac{1}{v_{ij}}\log\lp{\frac{ 1 + v_{ij} }{ 1 - v_{ij} }}\rp\,, \\
  \label{eq:IMM_general_1}
  I_{\mathrm{MM}}^{( 1)}(q_i,q_j) ={}& \frac{1-\scprod{\vec{u}}{\vec{w}}}{\sqrt{ A^2 - B^2 }} \lp{ K(z_2) - K(z_1) }\rp\,.
\end{align}
 
\paragraph{One massive and one massless emitter}
The reported formulae correspond to the result outlined in
Eqs.~(A.22) to (A.24) of Ref.~\cite{Alioli:2010xd}.
We consider one time-like momentum $q_i^{\,2} = m_i{\,^2}$ and one
light-like momentum $q_j^{\,2} = 0$. We define the symbol
\begin{align}
  \kappa ={} \sqrt{ 1 - \frac{ m_i^{\,2} }{ E_i^{\,2} } }\,.
\end{align}
Then the coefficients of the massive-massless angular integral are
%%%texparser:start:SijInt%%%
\begin{align}
  %%%texparser:LHS:IM0[-1]%%%
  I_{\mathrm{M0}}^{(-1)}(q_i,q_j)
  ={}&
       -1 \,, \\
  %%%texparser:LHS:IM0[0]%%%
  I_{\mathrm{M0}}^{( 0)}(q_i,q_j)
  ={}&
       \log\lp{\frac{ \scprod{\hat{q}_i}{\hat{q}_j}^2 }{ \scprod{\hat{q}_i}{\hat{q}_i} }}\rp\,, \\
  %%%texparser:LHS:IM0[1]%%%
  I_{\mathrm{M0}}^{( 1)}(q_i,q_j)
  ={}&
       -2\left\{
       \frac{1}{4}\log^2\lp{\frac{1-\kappa}{1+\kappa}}\rp
       + \log\lp{\frac{ \scprod{\hat{q}_i}{\hat{q}_j} }{1+\kappa}}\rp\log\lp{\frac{ \scprod{\hat{q}_i}{\hat{q}_j} }{1-\kappa}}\rp
       \right.
       \nonumber\\
     &\left.\hspace{1cm}
       + \mathrm{Li}_2\lp{ 1 - \frac{ \scprod{\hat{q}_i}{\hat{q}_j} }{1+\kappa} }\rp
       + \mathrm{Li}_2\lp{ 1 - \frac{ \scprod{\hat{q}_i}{\hat{q}_j} }{1-\kappa} }\rp
       \right\}
  \,.
\end{align}
%%%texparser:stop%%%

\paragraph{Two massive back-to-back emitters}
We consider the special case of two massive emitters with the same mass,
$q_i^{\,2} = m^2$ and $q_j^{\,2} = m^2$, arranged in a back-to-back
configuration, $\vec{q}_i = - \vec{q}_j$. We then have
$E_i = E_j = \mh/2$ and, therefore, $\kappa = \beta$. The expansion
coefficients are given by
%%%texparser:start:SijInt%%%
\begin{align}
  %%%texparser:LHS:IMMb2b[-1]%%%
  I_{\mathrm{MM},\mathrm{b2b}}^{(-1)}(q_i,q_j)
  ={}&
       0
       \,, \\
  %%%texparser:LHS:IMMb2b[0]%%%
  I_{\mathrm{MM},\mathrm{b2b}}^{( 0)}(q_i,q_j)
  ={}&
       -\frac{1+\beta^2}{\beta}\log\lp{\frac{1-\beta}{1+\beta}}\rp
       \,, \\
  %%%texparser:LHS:IMMb2b[1]%%%
  I_{\mathrm{MM},\mathrm{b2b}}^{( 1)}(q_i,q_j)
  ={}&
       \frac{1+\beta^2}{\beta}\left\{
        \mathrm{Li}_2\lp{ \frac{2\beta}{1+\beta}}\rp
       -\mathrm{Li}_2\lp{-\frac{2\beta}{1-\beta}}\rp
       \right\}
       \,, \\
  %%%texparser:LHS:IMMb2b[2]%%%
  I_{\mathrm{MM},\mathrm{b2b}}^{( 2)}(q_i,q_j)
  ={}&
       \frac{1+\beta^2}{\beta}\left\{
       -\frac{1}{3}\log^3\lp{\frac{1-\beta}{1+\beta}}\rp
       +2\mathrm{Li}_3\lp{ \frac{2\beta}{1+\beta}}\rp
       -2\mathrm{Li}_3\lp{-\frac{2\beta}{1-\beta}}\rp
       \right.
       \nonumber\\
     &
       \hspace{1cm}\left.
       -\log\lp{\frac{1-\beta}{1+\beta}}\rp\lb{
        \mathrm{Li}_2\lp{ \frac{2\beta}{1+\beta}}\rp
       +\mathrm{Li}_2\lp{-\frac{2\beta}{1-\beta}}\rp
       }\rb
       \right\}
       \,,
\end{align}
%%%texparser:stop%%%
where the subscript ``$\mathrm{b2b}$'' indicates a back-to-back
configuration of the emitters. Formulae for
$I_{\mathrm{MM},\mathrm{b2b}}^{(k)}$ for $k \in \{-1,0,1\}$ can be
obtained from \crefrange{eq:IMM_general_-1}{eq:IMM_general_1} by taking
the relevant limit. The formula for $I_{\mathrm{MM},\mathrm{b2b}}^{(2)}$
was calculated independently; this $\mathcal{O}(\ep^2)$ term is
needed in the soft limit of the real-virtual contribution.

\paragraph{Self-correlated massive emitter}
We consider the special case of a self-correlated massive emitter,
$i=j$, for a time-like momentum $q_i$ with $q_i^{\,2} = m^2$. Then we
have
%%%texparser:start:SijInt%%%
\begin{align}
  %%%texparser:LHS:IMMse[-1]%%%
  I_{\mathrm{MM}}^{(-1)}(q_i,q_i)
  ={}&
       0
       \,, \\
  %%%texparser:LHS:IMMse[0]%%%
  I_{\mathrm{MM}}^{( 0)}(q_i,q_i)
  ={}&
       2
       \,, \\
  %%%texparser:LHS:IMMse[1]%%%
  I_{\mathrm{MM}}^{( 1)}(q_i,q_i)
  ={}&
       -\frac{2}{\kappa}\log\lp{\frac{1-\kappa}{1+\kappa}}\rp
       \,, \\
  %%%texparser:LHS:IMMse[2]%%%
  I_{\mathrm{MM}}^{( 2)}(q_i,q_i)
  ={}&
       \frac{2}{\kappa}\left\{
        \mathrm{Li}_2\lp{ \frac{2\kappa}{1+\kappa}}\rp
       -\mathrm{Li}_2\lp{-\frac{2\kappa}{1-\kappa}}\rp
       \right\}
       \,.
\end{align}
%%%texparser:stop%%%
The $\order{\ep^2}$ term was calculated independently.

\subsection{Single-soft subtraction terms (one-loop)}
\label{sec:Rij1int}
We consider the soft limit of the one-loop \hbb{g} amplitude which is
given in \cref{eq:soft-fact-1loop}.
This factorisation involves both tree-level and one-loop soft
functions. The integration of the  tree-level eikonal factors is
discussed in \cref{sec:Sij0int}, while in this section we focus on the
integration of those terms in the one-loop soft function of
\cref{eq:soft-fact-1loop} which contain $\Rij{1}{ij}{4}$.

%%%
We integrate the one-loop eikonal factors over the soft-gluon phase
space, i.e.
\begin{align}
  \RInt{1}
  ={}&
            -\int \dq{4}
            \lp{ \Sij{0}{22}{4} -2 \Sij{0}{23}{4} + \Sij{0}{33}{4}}\rp
            \Rij{1}{23}{4}
            \,,
\end{align}
where $\Sij{0}{ij}{4}$ and $\Rij{1}{23}{4}$ are defined in
\cref{eq:single-eikonal,eq:Rij-def}, respectively. Since only
$\Rij{1}{23}{4}$ appears, we restrict ourselves to the case of two
emitters with the same non-vanishing mass in a back-to-back
configuration.

The soft-gluon phase space is parametrised using
\cref{eq:phsp_hbbg}. Note that even though the soft-gluon momentum
factorises from the energy-momentum conserving $\delta$-function, we
restrict ourselves to the same integration region as stated in
\cref{eq:phsp_hbbg}; this is in accordance with the choice made for
the integrated tree-level eikonal factor.
The only non-trivial integral is the integration over the angle $\theta$
between the momentum of the soft gluon $q_4$ and the $b$-quark momentum
$q_i$. We reexpress this angle in terms of the variable
$\lambda = \cos\theta$ and use the symmetry under $\lambda \to -\lambda$
to restrict the domain of integration to $\lambda \in [0,1]$. We arrive
at
\begin{align}
  \label{eq:SRij-ps}
  \int \dq{4}
    &= \frac{\muR^{2\ep} \Sep \Emax^{2-2\ep}}{(2 \pi)^{3-2\ep}}
       \int\dOmega{4}{2-2\ep} \int_0^1 \d\xi \, \xi^{1-2\ep}
       \int_0^1 \d\lambda \, (1-\lambda^2)^{-\ep}
  \,.
\end{align}
The energy dependence factorises from the integrand, which means that
the integral over $\d\xi$ can be solved trivially. The angular integral
$\dOmega{4}{2-2\ep}$ is performed using \cref{eq:dOmega_int}.

The expressions for the expansion coefficients $R_k(q_i,q_j;q_4)$
in Ref.~\cite{Czakon:2018iev} are given in terms of the variables
\begin{align}
  \alpha_i
    &= \frac{m_i^2}{2}
       \frac{\scprod{q_j}{q_4}}{\scprod{q_i}{q_4} \scprod{q_i}{q_j}}
     = \frac{(1-\beta^2) (1 + \beta \lambda)}{%
             2 (1+\beta^2) (1 - \beta \lambda)}
  \,, \\
  \alpha_j
    &= \frac{m_j^2}{2}
       \frac{\scprod{q_i}{q_4}}{\scprod{q_j}{q_4} \scprod{q_i}{q_j}}
     = \frac{(1-\beta^2) (1 - \beta \lambda)}{%
             2 (1+\beta^2) (1 + \beta \lambda)}
  \,,
\end{align}
and contain at most classical polylogarithms with arguments composed of
these variables. We rewrite those special functions in terms of
iterated integrals of argument $\lambda$ over the alphabet
\begin{align}
  &\frac{\d\lambda}{\lambda}, &
  &\frac{\d\lambda}{1+\lambda}, &
  &\frac{\d\lambda}{1-\lambda}, &
  &\frac{\d\lambda}{1+\beta \lambda}, &
  &\frac{\d\lambda}{1-\beta \lambda}\,.
\end{align}
As the rational coefficients in front of the iterated integrals also
only contain these letters, the integration over $\lambda$ can again be
performed in terms of iterated integrals over the same alphabet. These
iterated integrals are then evaluated at $1$ and we rewrite them in
terms of harmonic polylogarithms~\cite{Remiddi:1999ew} of argument
$\beta$ with up to weight four. All manipulations of the iterated
integrals are performed with the \Mathematica{} package
\HarmonicSums{}~\cite{Ablinger:2010kw,%
  Ablinger:2013hcp,Vermaseren:1998uu,Remiddi:1999ew,Blumlein:2009ta,%
  Ablinger:2011te,Ablinger:2013cf,Ablinger:2014rba,Ablinger:2016ll,%
  Ablinger:2017rad,Ablinger:2019mkx}.

Note that the term $\RInt{1}$ is again a Laurent series in
$\ep$. Therefore, we write
\begin{align}
  \label{eq:SRij-def}
  \RInt{1}
  ={}&
       \frac{ 4\CA }{ (4\pi)^2 }
       \lp{\frac{ \muR^2 }{ 4\Emax^2 }}\rp^{2\ep}
       \sum_{k=-3}^{0}
       \RIntCoeff{1}{k}(\beta) \, \ep^k
       \,.
\end{align}
The expansion coefficients $\RIntCoeff{1}{k}$ for $\mb=4.78~\gev$ and
$\mh=125.09~\gev$ evaluate to
\begin{align}
  \label{eq:SRintCoeff_numbers}
  \RIntCoeff{1}{-3}(\beta) ={}&  5.52628705137596 \,, \nonumber\\
  \RIntCoeff{1}{-2}(\beta) ={}&  35.3923534452863 \,, \nonumber\\
  \RIntCoeff{1}{-1}(\beta) ={}&  111.992677970445 \,, \nonumber\\
  \RIntCoeff{1}{ 0}(\beta) ={}&  245.654621810082 \,. 
\end{align}
The analytical expressions can be found in computer-readable form in an
ancillary file supplied with this article.

\subsection{Double-soft subtraction terms}
\label{sec:SSij1int}

%%% numerical double-soft treatment
For the integrated subtraction terms of the double-real contribution
we have to integrate the double-soft function, discussed in
\cref{sec:fact-double-soft}, over the unresolved phase space in $d$
dimensions. We define the integrated double-soft function as
\begin{align}
  \label{eq:dble_soft_ij_int_def}
  \DSoftInt{0}{ij}{q_2,q_3}
  ={}&
       \int \dq{4} \dq{5}\,
       \DSoft{0}{ij}(q_2,q_3;q_4,q_5)
  \,.
\end{align}

We use the phase-space parametrisation outlined in \cref{eq:phsp_hbbgg},
taking into account that the gluon momenta decouple from the overall
energy-momentum conserving $\delta$-function. Note that the decoupling
of soft particles means that, in principle, their energies are unbounded
unless we introduce some constraint by hand. For simplicity, we choose
to keep the bound that explicitly appears in the formulation of the
phase-space measure, i.e. we keep $E_{45,\mathrm{max}}$ unchanged
while the $\d\xi_1 \, \d\xi_2$ integration region still covers the unit
square.

It is particularly useful to keep the Born configuration fixed so that
the $b$-quark points along the $\hat{z}$-axis, i.e.
\begin{align}
  &q_2 ={} \tfrac{1}{2}\mh \lp{ \hat{t}^{\mu}  + \beta \hat{z}^{\mu} }\rp & \textrm{ and } &
  &q_3 ={} \tfrac{1}{2}\mh \lp{ \hat{t}^{\mu}  - \beta \hat{z}^{\mu} }\rp \,,
\end{align}
where $\hat{t}^{\mu}$ and $\hat{z}^{\mu}$ are unit vectors along time
and $\hat{z}$ axes, respectively.
After expressing the integrand using the phase-space parametrisation,
the $\xi_1$ dependence factorises and can be integrated analytically.
The remaining integrals are performed numerically.

The integrands of \cref{eq:dble_soft_ij_int_def} are still
divergent in the strongly-ordered (\soft{5}) and
collinear (\coll{45}) limits.
These divergences can be handled using the endpoint subtraction
method, as discussed in \cref{sec:philosophy}. We write
\begin{align}
  \label{eq:dble-soft-nested}
  \dsoft{45}
  ={}&
  (\ident{}-\soft{5})(\ident{}-\coll{45})\dsoft{45}
  + (\ident{}-\soft{5})\coll{45}\dsoft{45}
  + \soft{5}(\ident{}-\coll{45})\dsoft{45}
  + \soft{5}\coll{45}\dsoft{45}
  \,.
\end{align}
Note that the regularisation in the strongly-ordered limit applies only
to the $gg$ emission case of \cref{eq:dble_soft_gg_def}.
The relevant subtraction terms in \cref{eq:dble-soft-nested} are
constructed using the factorisation formulae from \cref{sec:fact-form}.

\begin{table}
\centering
\begin{tabular}{lllll}
\toprule
&
$k = -3$&
$k = -2$&
$k = -1$&
$k =  0$\\
\midrule
$C^{(k)}_{gg,\CA\CF}$&
$ -22.105148(3)$&
$ -120.8071(1)$&
$ -337.441(1)$&
$ -613.869(4)$\\
$C^{(k)}_{gg,\CF^2}$&
\multicolumn{1}{c}{--}&
$ +244.3194(4)$&
$ +915.818(2)$&
$ +984.741(6)$\\
$C^{(k)}_{\qq,\CF\TF}$&
\multicolumn{1}{c}{--}&
$ +7.368385(3)$&
$ +39.9006(1)$&
$ +124.966(1)$\\
\bottomrule
\end{tabular}
\caption{Coefficients of the $\ep$ expansion of the integrated
  double-soft function for $gg$ channel (first two rows), and the $\qq$
  channel (the last row). We use $\mb=4.78~\gev$ and $\mh=125.09~\gev$.}
\label{tab:dble_soft_gg_int}
\end{table}
The integrated double-soft function can be written as a Laurent series,
i.e.
\begin{align}
  \label{eq:dble_soft_ij_int}
  \DSoftInt{0}{ij}{q_2,q_3}
  ={}&
       \frac{1}{(4\pi)^4}
       \,
       \lp{\frac{\muR^2}{\Emax^2}}\rp^{2\ep}
       \sum_{k=-3}^0 C^{(k)}_{ij} \ep^k
  \,.
\end{align}
Note that the highest pole that occurs in the $gg$ channel is
$\ep^{-3}$. This arises from taking all singular limits in the
integrand, i.e. the strongly-ordered, collinear and the double-soft
limits. In the $\qq{}$ case we only have an $\ep^{-2}$ pole since the
strongly-ordered soft limit, $\soft{5}$, is regular.
Furthermore, we split the $C^{(k)}_{ij}$ coefficients into separate
colour structures,
\begin{align}
  C^{(k)}_{gg}  ={}& \CA\CF \, C^{(k)}_{gg,\CA\CF} + \CF^{\,2} \, C^{(k)}_{gg,\CF^2} \,, \\
  C^{(k)}_{\qq} ={}& \CF\TF \, C^{(k)}_{\qq,\CF\TF}\,.
\end{align}
%% results
We present numerical values for the integrated double-soft function
coefficients, evaluated for $\mb=4.78~\gev$ and $\mh=125.09~\gev$, in
\cref{tab:dble_soft_gg_int}.

% -------------------------
\bibliographystyle{utphys}
\bibliography{h2bb}{}
\end{document}